\documentclass[pdflatex,sn-basic]{sn-jnl}

\usepackage{graphicx}
\usepackage{multirow}
\usepackage{amsmath,amssymb,amsfonts}
\usepackage{helvet}
\usepackage{xcolor}
\definecolor{snblue}{HTML}{2563EB}
\definecolor{snteal}{HTML}{0EA5A4}
\definecolor{snamber}{HTML}{F59E0B}
\definecolor{sngrey}{HTML}{64748B}
\usepackage{textcomp}
\usepackage{booktabs}
\usepackage{listings}
\usepackage{enumitem}
\usepackage{siunitx}
\usepackage{xspace}
\usepackage{url}
\usepackage{placeins}
\usepackage{tikz}
\usetikzlibrary{shapes.geometric, arrows.meta, positioning, fit, calc, backgrounds}

\newcommand{\subtool}{\textsc{cukereuse-subscenarios}\xspace}
\newcommand{\cukereuse}{\textsc{cukereuse}\xspace}

\AtBeginDocument{\hypersetup{bookmarksdepth=3}}

\begin{document}

\title[Given, When, Then, Again: Subscenario Refactoring Candidates in BDD Suites]{Given, When, Then, Again: Mining Subscenario Refactoring Candidates in Behaviour-Driven Test Suites with ML Classifiers and LLM-Judge Baselines}

\author*[1]{\fnm{Ali Hassaan} \sur{Mughal}}\email{alihassaanmughal.work@gmail.com}

\author[2]{\fnm{Noor} \sur{Fatima}}\email{nfatima.bce25seecs@seecs.edu.pk}

\author[3]{\fnm{Muhammad} \sur{Bilal}}\email{m.bilal@tum.de}

\affil*[1]{\orgname{Independent Researcher; Applied MBA (Data Analytics), Texas Wesleyan University}, \orgaddress{\city{Fort Worth}, \state{TX}, \country{USA}}}

\affil[2]{\orgname{Independent Researcher; B.E.\ Computer Engineering, National University of Sciences and Technology (NUST)}, \orgaddress{\country{Pakistan}}}

\affil[3]{\orgname{Independent Researcher; M.Sc.\ Management, Technical University of Munich}, \orgaddress{\city{Munich}, \country{Germany}}}

\abstract{\textbf{Context.} Behaviour-Driven Development (BDD) software test suites accumulate duplicated step subsequences. Three published refactoring patterns are available (within-file Background, within-repo reusable-scenario invocation, cross-organisational shared higher-level step), but no prior work automates which recurring subsequences are worth extracting or which mechanism applies.

\textbf{Objective.} Rank recurring step subsequences (``slices'') by refactoring suitability (extraction-worthy), pre-map each to one of the three patterns, and quantify prevalence across the public BDD ecosystem.

\textbf{Method.} Every contiguous $L$-step window ($L \in [2, 18]$) in a 339-repository / 276-upstream-owner Gherkin corpus is keyed by paraphrase-robust cluster identifiers and counted under three scopes. Sentence-BERT (SBERT) / Uniform Manifold Approximation and Projection (UMAP) / Hierarchical Density-Based Spatial Clustering of Applications with Noise (HDBSCAN) recovers paraphrase-equivalent slices. Three authors label a stratified 200-slice pool against a written rubric. An eXtreme Gradient Boosting (XGBoost) extraction-worthy classifier trained under 5-fold cross-validation is compared with a tuned rule baseline and two open-weight Large Language Model (LLM) judges.

\textbf{Results.} The miner produces 5{,}382{,}249 slices collapsing to 692{,}020 recurring patterns. Three-author Fleiss' $\kappa = 0.56$ (extraction-worthy) and $0.79$ (mechanism). The classifier reaches out-of-fold $F_1 = 0.891$ (95\,\% CI $[0.852, 0.927]$), outperforming both the rule baseline ($F_1 = 0.836$, $p = 0.017$) and the better LLM judge ($F_1 = 0.728$, $p = 1.5\times10^{-4}$). 75.0\,\%, 59.5\,\%, and 11.7\,\% of scenarios carry a within-file Background, within-repo reusable-scenario, and cross-organisational shared-step candidate, respectively; the figures are stable under a sweep of the classifier decision threshold.

\textbf{Conclusion.} Paraphrase-robust subscenario discovery yields a corpus-wide census of BDD refactoring candidates; pipeline, classifier predictions, labelled pool, and rubric are released under Apache-2.0.}


\maketitle

\section{Introduction}
\label{sec:intro}

\paragraph{What this paper adds.}
We provide, to our knowledge, the first \emph{paraphrase-robust} subsequence miner for
BDD software-test specifications: a static pipeline that turns a
1.1M-step corpus of Gherkin tests into a ranked list of
\emph{contiguous step-subsequences}
that are worth extracting, each pre-mapped to one of three concrete
reuse mechanisms with a published Cucumber-Java implementation
\citep{mughal2024bdd}. Prior BDD work either operates at whole-scenario
granularity \citep{binamungu2018saner,binamungu2020xp,oliveira2019quality} or
studies refactoring support without a corpus-scale miner
\citep{irshad2022ist,irshad2021jss}; sequence-mining
classics (PrefixSpan, SPADE) work on exact symbol sequences and do not
handle the surface-paraphrase variation BDD steps exhibit.
Table~\ref{tab:novelty-positioning} positions the contribution against
the closest prior work along five axes. Three deliverables make the
paper concrete and reproducible: (i)~a slice inventory of
$5{,}382{,}249$ contiguous $L$-step windows ($L \in [2,18]$)
collapsing to $692{,}020$ recurring cluster-id-sequence patterns
across the $339$-repo slice-bearing subset of the $347$-repo
\cukereuse corpus, spanning $276$ distinct upstream owners on GitHub
(a mix of Organisation and User accounts; see
Section~\ref{sec:limitations}), of which $30{,}955$ recur across
$\geq 2$ distinct upstream owners;
(ii)~a stratified three-author labelling protocol ($n=200$,
$60$-slice overlap, four-category Fleiss'
$\kappa = 0.560$) extending the pair-level \cukereuse rubric to
slice level; and (iii)~an eXtreme Gradient Boosting (XGBoost)
extraction-worthy classifier that reaches out-of-fold $F_1 = 0.891$
($95\,\%$ bootstrap CI $[0.852, 0.927]$) and outperforms two
open-weight Large Language Model (LLM) judges ($F_1 \leq 0.728$)
on the same task.

BDD software-test suites written in Gherkin accumulate the same
maintenance hazards as unit-test corpora: copy-paste setup,
parameter-varying near-duplicates, and silent drift. Our predecessor \cukereuse
documents the scale at the step level: across 347 public GitHub
repos and 1{,}113{,}616 parsed Gherkin steps, 80.2\,\% of step
occurrences are byte-identical copies seen elsewhere, markedly
higher than is typical of production source code
\citep{baxter1998clonedr,roy2009clonesurvey}.
The unit of practical refactoring, however, is rarely an isolated
step; it is a contiguous \emph{run} of two or more steps
(a \emph{slice}). \citet{mughal2024bdd} supplies a Cucumber-Java
implementation of three reuse mechanisms (\texttt{Background}, a
reusable scenario invoked via a single-step call, and a custom
higher-level step from a two-stage code generator);
\citet{mughal2026cukereuse} supplies the corpus and the per-step
paraphrase-robust cluster identifier; the present paper supplies
the \textbf{discovery layer} that automatically identifies which
slices to extract and which of the three mechanisms applies.

\paragraph{Research questions.}
We organise the analysis around three research questions, each tied
to one Mughal-2024 mechanism:
\textbf{RQ1} (within-file): how prevalent are step subsequences
that recur across scenarios in the same \texttt{.feature} file
(\texttt{Background}-block candidates)?
\textbf{RQ2} (within-repo cross-file): how prevalent are
subsequences shared across \texttt{.feature} files of one repository
(reusable-\texttt{.feature} candidates, invoked via
\texttt{I~call~feature~\allowbreak file~$\langle\,X\rangle$})?
\textbf{RQ3} (cross-organisational): how prevalent are
subsequences that paraphrase-cluster across repositories owned by
different upstream owners (custom higher-level-step candidates)?

\paragraph{Contributions.}
(1)~a paraphrase-robust subsequence miner keyed by \cukereuse
hybrid cluster identifiers, with a Sentence-BERT (SBERT) /
Uniform Manifold Approximation and Projection (UMAP) /
Hierarchical Density-Based Spatial Clustering of Applications
with Noise (HDBSCAN) slice-embedding pass that recovers
paraphrase-equivalent slices missed by exact cluster-id matching;
(2)~a three-scope ranking (within-file, within-repo, cross-org)
that distinguishes the three Mughal-2024 mechanisms;
(3)~a stratified three-author labelling protocol on a 200-slice
pool with a 60-slice overlap subset, extending the \cukereuse
pair-level methodology to slice level;
(4)~a two-stage XGBoost classifier (binary extraction-worthy +
three-way mechanism), benchmarked head-to-head against a tuned rule
baseline (McNemar $\chi^2 = 5.69$, $p = 0.017$) and against two
open-weight LLM-judge baselines (\texttt{openai/gpt-oss-120b},
\texttt{inclusionai/ling-2.6-1t}; McNemar $\chi^2 \geq 14.4$,
$p \leq 1.5\times10^{-4}$) on the same human-anchored pool, with all classifier
predictions, per-judge raw outputs, labels, and rubric released under
Apache-2.0.

\begin{table*}[!ht]
\centering
\caption{Positioning against closest prior work. Columns:
\emph{Granularity} = duplication unit; \emph{Mode} = static / dynamic
/ survey; \emph{Scale} = largest corpus applied; \emph{Para.-robust} =
tolerates step-text paraphrase; \emph{Mech-mapped} = candidate
pre-mapped to a refactoring mechanism with published implementation.}
\label{tab:novelty-positioning}
\footnotesize
\setlength{\tabcolsep}{4pt}
\resizebox{\linewidth}{!}{%
\begin{tabular}{lllrcc}
\toprule
Work & Granularity & Mode & Scale & Para.-robust & Mech-mapped \\
\midrule
\citet{binamungu2018saner,binamungu2020xp}                              & whole scenario  & dynamic    & $\sim$1 repo   & no  & no  \\
\citet{oliveira2017quality,oliveira2019quality}                         & whole scenario  & catalogue  & manual         & no  & no  \\
\citet{irshad2022ist,irshad2021jss}                                     & spec-level      & industrial & 2 projects     & --- & no  \\
GivenWhenThen \citep{alcantara2026gwt}                                  & step + step-def & static     & 1{,}720 repos  & no  & no  \\
PrefixSpan / SPADE \citep{pei2001prefixspan,zaki2001spade}              & symbol sequence & static     & ---            & no  & --- \\
\cukereuse \citep{mughal2026cukereuse}                                  & single step     & static     & 347 repos      & yes & no  \\
\textbf{This paper (\subtool)}                                          & \textbf{contiguous slice} & \textbf{static} & \textbf{339 / 347 repos} & \textbf{yes} & \textbf{yes (3)} \\
\bottomrule
\end{tabular}}
\end{table*}

\FloatBarrier
\section{Background and motivation}
\label{sec:background}

We inherit two assets and target one gap. From \cukereuse
\citep{mughal2026cukereuse}: a 347-repo Gherkin corpus and a
paraphrase-robust hybrid \texttt{cluster\_id} per step (three-author
Fleiss' $\kappa = 0.84$ on a 1{,}020-pair benchmark). Re-keying the
steps of a slice by their \texttt{cluster\_id}s gives slice identity
that is robust to surface paraphrasing across repositories and
framework dialects. From \citet{mughal2024bdd}: three Cucumber-Java
reuse mechanisms with a published implementation
(Section~\ref{sec:approach:mapping}): a \texttt{Background:}
block (within-file), a reusable scenario invoked via
\texttt{I~call~feature~\allowbreak file~$\langle\,X\rangle$} (within-repo), and
a shared higher-level step generated from a Java \texttt{enum}
walk of \texttt{features/} (cross-org). Section~6.4 of that paper
notes that the \emph{discovery} of which slices are worth extracting
is left to manual review, and that the manual cost grows
super-linearly with suite size. That is the gap this paper closes.
\cukereuse measures duplication at the single-step granularity;
the practical refactoring unit, however, is the contiguous
\emph{run} of two or more consecutive steps, since each of the
three mechanisms above operates on a multi-step block. The discovery
question is therefore not \emph{which steps recur} but \emph{which
contiguous step subsequences recur in a way that maps onto an
extraction mechanism}.

\FloatBarrier
\section{Related work}
\label{sec:related}

\subsection{BDD scenario quality, smells, and refactoring}

BDD was scoped as an engineering practice by \citet{north2006bdd}
and characterised in agile-acceptance-testing use by
\citet{solis2011bdd}. The mapping study of \citet{binamungu2023jss}
and the field study of \citet{pereira2018bdd} converge on the same
pain points: duplicated setup, brittle assertions, opaque scenario
boundaries; \citet{scandaroli2019bdd} report two industrial cases
where this maintenance burden dominates steady-state cost. A
parallel quality-rubric line
\citep{oliveira2017quality,oliveira2019quality,wautelet2023poem,sears2025profes}
operationalises ``a good BDD scenario'' at the whole-scenario
unit; the rubrics help an author judge a scenario but do not
identify which sub-sequences recur and warrant extraction, the
question we address.

The closest empirical work is the Binamungu et al.\ trio
\citep{binamungu2018vst,binamungu2018saner,binamungu2020xp},
which detects duplicate \emph{whole scenarios} \emph{dynamically}
(by comparing executed step traces) on a small handful of
repositories. We are static, operate at contiguous-subsequence
granularity, and span more than an order of magnitude more repositories.
\citet{irshad2022ist,irshad2021jss} study the refactoring and
large-scale adoption of BDD specifications in industrial settings
and document target \emph{identification} as the limiting cost. We
automate that identification step at corpus scale.

A concurrent BDD dataset, GivenWhenThen \citep{alcantara2026gwt},
was released in the same cycle as \cukereuse on a disjoint
1{,}720-repo sample with a different granularity (each scenario
paired with its backing step-definition source); nothing in our
pipeline depends on GivenWhenThen (GWT), but the slice-mining methodology
generalises naturally to it.

\subsection{Sequence mining}

Frequent-pattern mining over sequence databases has been studied
since \citet{agrawal1995sequential}; GSP \citep{srikant1996gsp},
PrefixSpan \citep{pei2001prefixspan}, and SPADE \citep{zaki2001spade}
introduce the canonical pattern-growth and vertical id-list
formulations, with the closely related discovery of partially
ordered episodes covered by \citet{mannila1997episodes}. The
surveys of \citet{fournierviger2017survey} and \citet{bechini2023spm}
catalogue the exact / approximate / gap-constrained / closed
trade-offs; the closed-pattern formulation underlies our R6
closure filter (Section~\ref{sec:threats}).

We use exact $n$-gram counting on cluster-id
sequences with $L \in [2, 18]$ rather than PrefixSpan or SPADE in
canonical form: the per-scenario search space is small, the
cluster-id alphabet is finite, and exact counting is sufficient,
parallelisable, and trivially correct. PrefixSpan with
\texttt{gap}~$\leq 1$ is used (Phase~3) only as a robustness check
for slices interrupted by one intervening step.

\subsection{Code-clone detection}

The clone-detection literature supplies the algorithmic
heritage. Abstract Syntax Tree (AST) and token-level clone detectors
\citep{baxter1998clonedr,kamiya2002ccfinder,li2004cpminer,jiang2007deckard}
matured into corpus-scale tools
\citep{sajnani2016sourcerercc,saini2018oreo}, evaluated against the
Bellon benchmark \citep{bellon2007clones} and more recently the
\citet{krinke2025bigclonebench} re-labelling that exposes
weak-Type-3/Type-4 mis-labels in BigCloneBench. The Roy--Cordy
taxonomy
\citep{roy2009clonesurvey,rattan2013clonesurvey}
classifies clones from Type~1 (textual identity) through Type~4
(functional equivalence). In that taxonomy our exact cluster-id
sequence match is Type~1 (cluster-ids are the token alphabet);
the Phase~4 SBERT/UMAP/HDBSCAN clustering recovers Type~3/4
paraphrase equivalence by collapsing slices whose texts are
semantically near-equivalent. We re-use the vocabulary; we do not
re-implement clone detection.

\subsection{Software-test-suite refactoring and minimisation}

\citet{yoo2012regression} survey regression-test minimisation,
selection, and prioritisation; the \emph{subsumption} /
\emph{redundancy} / \emph{coverage-equivalence} vocabulary frames
our recommendations as coverage-preserving refactoring.
Software-test-smell catalogues
\citep{garousi2018smells,bavota2012testsmells,panichella2022testsmells20}
identify duplicated setup as a maintainability hazard
(the unit-test analogue of our RQ1 within-file recurrence),
and \citet{pontillo2024mltestsmell} extend the catalogue with an
ML-based detector, a parallel to our extraction-worthy classifier.
Recent software-test refactoring evidence
\citep{martins2025testref,martins2024catalog,horikawa2025testref,liu2025llmrefactor}
agrees that an extraction gate plus a concrete mechanism mapping
is the right intervention shape (the last finding: even strong LLM
refactoring agents struggle without explicit refactoring-type
guidance).

The closest non-BDD analogue is the test-clone literature, which
treats each test case as indivisible; we are not aware of a clone
study at contiguous sub-sequences \emph{within} test cases. Our
slice formulation makes that granularity tractable for BDD
specifically by lifting identity from raw text to the \cukereuse
paraphrase-robust cluster id.

\FloatBarrier
\section{Approach}
\label{sec:approach}

\subsection{Slice as unit of analysis}
\label{sec:approach:slice}

Let a \emph{scenario} be a sequence of $n$ Gherkin steps
$s_1, s_2, \ldots, s_n$, each step parsed by the \cukereuse pipeline
into a record carrying \texttt{(repo\_slug, file\_path, scenario,
keyword, text, cluster\_id, is\_background, is\_outline)}. A
\emph{slice} of length $L \in [2, L_{\max}]$ at position $p$ is the
sub-sequence $\langle\,s_p, s_{p+1}, \ldots, s_{p+L-1}\rangle$. Each
slice carries a \emph{cluster-id sequence}
$\langle\,c_p, c_{p+1}, \ldots, c_{p+L-1}\rangle$, where $c_i$ is the
\cukereuse hybrid \texttt{cluster\_id} of $s_i$. Two slices that
share the same cluster-id sequence count as the same logical slice
even when their underlying step text differs; this is our
\emph{paraphrase-robust slice identity}.

We restrict the mining to scenarios with $\geq 2$ steps remaining
after dropping rows where \texttt{is\_background = True} or where
the step has no assigned \texttt{cluster\_id}. The corpus contains
136{,}970 scenarios under the canonical key
\texttt{(repo\_slug, file\_path, scenario)} after dropping
\texttt{is\_background = True} rows (the same key paper~1 uses
implicitly via the \cukereuse parser's
\texttt{is\_background} field). After further filtering for empty
scenario names (Karate-style \texttt{*}-only files) and
length~$<2$ slices, the mining set is 134{,}635 scenarios.
Pre-flight (Phase~0; Section~\ref{sec:method}) sets
$L_{\max} = 18$, the 95th percentile of cleaned scenario lengths.

\subsection{Three scopes}
\label{sec:approach:scopes}

A slice's recurrence is interesting at three nested scopes
(Figure~\ref{fig:scopes}), each mapping to a different Mughal-2024
mechanism:

\begin{figure}[!ht]
\centering
\resizebox{\linewidth}{!}{\sffamily%
\begin{tikzpicture}[
  every node/.style={font=\footnotesize},
  scope3/.style={rectangle, rounded corners=4pt, draw=snamber!85!black,
                 fill=snamber!16, line width=0.7pt,
                 minimum width=70mm, minimum height=42mm},
  scope2/.style={rectangle, rounded corners=3pt, draw=snteal!85!black,
                 fill=snteal!15, line width=0.7pt,
                 minimum width=54mm, minimum height=30mm},
  scope1/.style={rectangle, rounded corners=2pt, draw=snblue!85!black,
                 fill=snblue!15, line width=0.7pt,
                 minimum width=38mm, minimum height=18mm},
  ftxt/.style={font=\scriptsize\bfseries},
  ntxt/.style={font=\scriptsize\itshape},
]
\node[scope3]                               (s3) {};
\node[ftxt, anchor=south west, text=snamber!75!black]
      at (s3.south west) {\quad RQ3: cross-organisational};
\node[ntxt, anchor=north west, text=snamber!75!black]
      at (s3.north west) {\quad orgs $A,B,C,\dots$};

\node[scope2, anchor=center]                (s2) at ($(s3.center)+(0,1mm)$) {};
\node[ftxt, anchor=south west, text=snteal!70!black]
      at (s2.south west) {\, RQ2: within-repo cross-file};
\node[ntxt, anchor=north west, text=snteal!70!black]
      at (s2.north west) {\, repo $r$};

\node[scope1, anchor=center]                (s1) at ($(s2.center)+(0,1mm)$) {};
\node[ftxt, anchor=south west, text=snblue!75!black]
      at (s1.south west) {RQ1: within-file};
\node[ntxt, anchor=north west, text=snblue!75!black]
      at (s1.north west) {file $f$};

\node[font=\scriptsize, anchor=center] at (s1.center)
      {slice $\sigma$};

\node[font=\scriptsize, anchor=west]
      at ([xshift=4mm]s3.east|-s3.north)
      {{\color{snamber!80!black}distinct orgs\,$\geq\!2$}};
\node[font=\scriptsize, anchor=west]
      at ([xshift=4mm]s3.east|-s2.north)
      {{\color{snteal!75!black}max within-repo files\,$\geq\!2$}};
\node[font=\scriptsize, anchor=west]
      at ([xshift=4mm]s3.east|-s1.north)
      {{\color{snblue!80!black}max within-file recurrence\,$\geq\!2$}};
\end{tikzpicture}}
\caption{Three nested scopes for slice recurrence (RQ1
within-file $\subset$ RQ2 within-repo $\subset$ RQ3 cross-org); a
single slice can qualify at multiple scopes.}
\label{fig:scopes}
\end{figure}
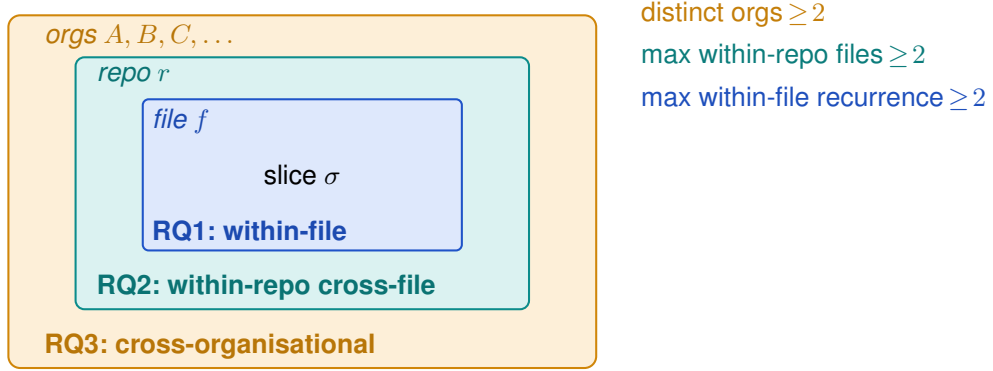

\begin{itemize}
\item \textbf{Within-file (RQ1).} Recurrence across scenarios in the
  same \texttt{.feature}: candidate for a top-of-file
  \texttt{Background} block (Mughal 2024 Section~1). Metric:
  \texttt{max\_within\_file\_recurrence}, the maximum over
  \texttt{(repo, file)} pairs of distinct scenarios containing the
  slice.

\item \textbf{Within-repo cross-file (RQ2).} Recurrence across files
  in one repository: candidate for extraction to a reusable
  \texttt{.feature} invoked via
  \texttt{I~call~feature~\allowbreak file~$\langle\,$ENUM$\rangle$} (Mughal 2024
  Section~4.1). Metric: \texttt{max\_within\_repo\_files}, the maximum
  over repositories of distinct containing files.

\item \textbf{Cross-organisational (RQ3).} Recurrence across repos
  owned by different upstream owners: candidate for promotion to a
  custom higher-level step backed by an Algorithm~2 step-definition
  method from \citet{mughal2024bdd}. Metric: \texttt{n\_distinct\_orgs}, the count
  of distinct upstream owners (segment before the first underscore in
  \texttt{repo\_slug}, equivalent to the top-level GitHub
  account-owner namespace; on GitHub this can be either an
  Organisation account or a User account, and the namespace boundary
  is what matters for cross-context recurrence; see
  Section~\ref{sec:limitations}). This is distinct from
  \texttt{n\_distinct\_repos}: a single owner publishing many
  repos (e.g., multi-language software-development-kit (SDK)
  clients) inflates the cross-repo
  signal without genuine cross-owner reuse (magnitude in
  Section~\ref{sec:discussion}).
\end{itemize}

\subsection{Mechanism mapping}
\label{sec:approach:mapping}

Each recurring slice is mapped to one of four conceptual targets;
the Phase-6 binary classifier handles the \texttt{no\_op} case and
Phase~8 then assigns one of the three concrete mechanisms to surviving
candidates. The four targets are:

\begin{enumerate}
\item \texttt{background}: prepend the slice to the file's
  \texttt{Background} block.
\item \texttt{reusable\_scenario}: emit a new \texttt{.feature}
  under \texttt{features/\allowbreak reusable/}\allowbreak$\langle\textit{group}\rangle$\allowbreak\texttt{/}\allowbreak$\langle\textit{name}\rangle$\allowbreak\texttt{.feature},
  insert \texttt{And~I~call~feature~file~}$\langle$\textit{ENUM}$\rangle$
  at each call site, and regenerate the ENUM constants via
  Algorithm~1 of \citet{mughal2024bdd}.
\item \texttt{shared\_higher\_level\_step}: promote the slice to
  a single named step backed by an Algorithm~2 step-definition method
  from \citet{mughal2024bdd}.
\item \texttt{no\_op}: the slice is not a useful extraction
  target, despite recurring.
\end{enumerate}

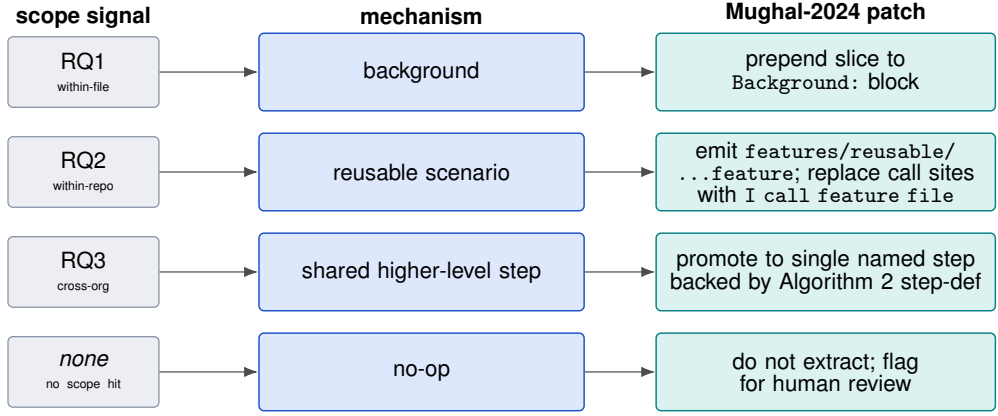
\begin{figure}[!ht]
\centering
\resizebox{\linewidth}{!}{\sffamily%
\begin{tikzpicture}[
  every node/.style={font=\scriptsize},
  scope/.style={rectangle, rounded corners=2pt, draw=sngrey!80,
                fill=sngrey!12, line width=0.5pt,
                text width=20mm, align=center, minimum height=10mm,
                inner sep=2pt},
  mech/.style={rectangle, rounded corners=2pt, draw=snblue!85!black,
                fill=snblue!15,    line width=0.6pt,
                text width=44mm, align=center, minimum height=11mm,
                inner sep=3pt},
  patch/.style={rectangle, rounded corners=2pt, draw=snteal!80!black,
                fill=snteal!15,   line width=0.6pt,
                text width=46mm, align=center, minimum height=11mm,
                inner sep=3pt},
  flow/.style={-{Latex[length=2mm]}, line width=0.5pt, draw=black!70},
]
\node[scope]                            (sf) {RQ1\\\tiny within-file};
\node[scope, below=4mm of sf]           (sr) {RQ2\\\tiny within-repo};
\node[scope, below=4mm of sr]           (so) {RQ3\\\tiny cross-org};
\node[scope, below=4mm of so]           (sn) {\textit{none}\\\tiny no scope hit};

\node[mech, anchor=west] (mb) at ([xshift=14mm]sf.east) {background};
\node[mech, anchor=west] (mr) at ([xshift=14mm]sr.east) {reusable scenario};
\node[mech, anchor=west] (ms) at ([xshift=14mm]so.east) {shared higher-level step};
\node[mech, anchor=west] (mn) at ([xshift=14mm]sn.east) {no-op};

\node[patch, anchor=west] (pb) at ([xshift=10mm]mb.east) {prepend slice to
  \texttt{Background:} block};
\node[patch, anchor=west] (pr) at ([xshift=10mm]mr.east) {emit \texttt{features/reusable/}\\
  \texttt{...feature}; replace call sites with
  \texttt{I call feature file}};
\node[patch, anchor=west] (ps) at ([xshift=10mm]ms.east) {promote to single named step
  backed by Algorithm~2 step-def};
\node[patch, anchor=west] (pn) at ([xshift=10mm]mn.east) {do not extract; flag for human
  review};

\foreach \s/\m in {sf/mb, sr/mr, so/ms, sn/mn}
  \draw[flow] (\s.east) -- (\m.west);
\foreach \m/\p in {mb/pb, mr/pr, ms/ps, mn/pn}
  \draw[flow] (\m.east) -- (\p.west);

\node[font=\scriptsize\bfseries, anchor=south] at (sf.north) {scope signal};
\node[font=\scriptsize\bfseries, anchor=south] at (mb.north) {mechanism};
\node[font=\scriptsize\bfseries, anchor=south] at (pb.north) {Mughal-2024 patch};
\end{tikzpicture}}
\caption{Mechanism mapping: scope signal $\to$ mechanism $\to$
concrete Mughal-2024 patch shape. Phase~6 gates whether the mapping
fires; Phase~8 refines the mechanism choice.}
\label{fig:mechmap}
\end{figure}

A scope-driven rule-based predictor (RQ1 $\to$ \texttt{background},
RQ2 $\to$ \texttt{reusable\_scenario},
RQ3 $\to$ \texttt{shared\_higher\_level\_step},
otherwise \texttt{no\_op}) provides a baseline; a learned classifier on the
labelled pool refines it. Figure~\ref{fig:mechmap} shows the
end-to-end mapping from scope signal to the concrete Mughal-2024
patch it implies.

\FloatBarrier
\section{Method}
\label{sec:method}

\begin{figure}[!ht]
\centering
\resizebox{0.92\linewidth}{!}{\sffamily%
\begin{tikzpicture}[
  node distance=2mm and 3mm,
  every node/.style={font=\scriptsize},
  auto/.style={rectangle, rounded corners=2pt, draw=snblue!85!black,
               fill=snblue!15,    line width=0.6pt, minimum height=7mm,
               text width=38mm, align=center, inner sep=2pt},
  human/.style={rectangle, rounded corners=2pt, draw=snamber!85!black,
               fill=snamber!18, line width=0.6pt, minimum height=7mm,
               text width=38mm, align=center, inner sep=2pt},
  rollup/.style={rectangle, rounded corners=2pt, draw=snteal!80!black,
               fill=snteal!15,   line width=0.6pt, minimum height=7mm,
               text width=38mm, align=center, inner sep=2pt},
  hdr/.style={font=\bfseries\footnotesize},
  flow/.style={-{Latex[length=2mm]}, line width=0.45pt, draw=black!70},
]
\node[hdr, anchor=south] (hA) at (0,0) {A. Mining};
\node[auto, below=2mm of hA]            (a0) {\textbf{Phase 0} corpus audit\\
  \scriptsize 134{,}635 scenarios, $L_{\max}\!=\!18$};
\node[auto, below=2mm of a0]            (a1) {\textbf{Phase 1} slice extraction\\
  \scriptsize 5{,}382{,}249 slices};
\node[auto, below=2mm of a1]            (a2) {\textbf{Phase 2 + 2c} $n$-gram
  + spec-suite detector\\
  \scriptsize 692{,}020 recurring patterns};
\node[auto, below=2mm of a2]            (a4) {\textbf{Phase 4} SBERT $\to$
  UMAP $\to$ HDBSCAN\\
  \scriptsize 33{,}121 paraphrase clusters};
\node[rollup, below=2mm of a4]          (a9) {\textbf{Phase 9a}
  pre-classifier prevalence};

\node[hdr, anchor=south] (hB) at (46mm,0) {B. Labelling (human)};
\node[auto, below=2mm of hB, dashed]    (b1) {\textbf{Pilot} 10-slice
  4-rater calibration\\
  \scriptsize informs Phase~2c};
\node[human, below=2mm of b1]           (b2) {\textbf{Rubric} written\\
  \scriptsize B-1..B-5 / N-1..N-5 / spec};
\node[human, below=2mm of b2]           (b3) {\textbf{Sampling} stratified
  on $L\!\times\!$scope$\!\times\!$support\\
  \scriptsize $n\!=\!200$, 60-overlap};
\node[human, below=2mm of b3]           (b4) {\textbf{Phase 5} 3-author labels\\
  \scriptsize Fleiss $\kappa\!=\!0.560$ (4-cat)\\
  \scriptsize Fleiss $\kappa\!=\!0.788$ (mech)};

\node[hdr, anchor=south] (hC) at (92mm,0) {C. Modelling};
\node[auto, below=2mm of hC]            (c6) {\textbf{Phase 6}
  XGBoost EW classifier\\
  \scriptsize $F_1\!=\!0.891$ {\scriptsize [0.852, 0.927]}};
\node[auto, below=2mm of c6]            (c7) {\textbf{Phase 7}
  LLM-judge baseline\\
  \scriptsize $F_1\!\leq\!0.728$ (gpt-oss-120b)};
\node[auto, below=2mm of c7]            (c8) {\textbf{Phase 8}
  mechanism predictor\\
  \scriptsize accuracy $=$ 0.965};
\node[rollup, below=2mm of c8]          (c9) {\textbf{Phase 9b}
  post-classifier headline\\
  \scriptsize 83.2\,\% / 43.7\,\% repos hit};

\foreach \x/\y in {a0/a1, a1/a2, a2/a4, a4/a9}
  \draw[flow] (\x) -- (\y);
\foreach \x/\y in {b1/b2, b2/b3, b3/b4}
  \draw[flow] (\x) -- (\y);
\foreach \x/\y in {c6/c7, c7/c8, c8/c9}
  \draw[flow] (\x) -- (\y);

\draw[flow, dashed] (a2.east) to[out=0,in=180] (b3.west);
\coordinate (gAB) at ([xshift=3mm]a4.east);     
\coordinate (gBC) at ([xshift=2.5mm]b4.east);   
\coordinate (bus) at ([yshift=-4mm]a9.south);   
\draw[dashed, line width=0.45pt, draw=black!70, rounded corners=1.5mm]
  (a4.east) -- (gAB) -- (gAB |- bus) -- (gBC |- bus) -- (gBC);
\draw[flow, dashed, rounded corners=1.5mm]
  (b4.east) -- (gBC) |- (c6.west);
\end{tikzpicture}}
\caption{Pipeline overview: \textbf{(A)} mining $\to$
\textbf{(B)} three-author labelling (the human bottleneck) $\to$
\textbf{(C)} classifier and mechanism predictor. Solid arrows are
within-column flow; dashed arrows are cross-column hand-offs.}
\label{fig:pipeline}
\end{figure}
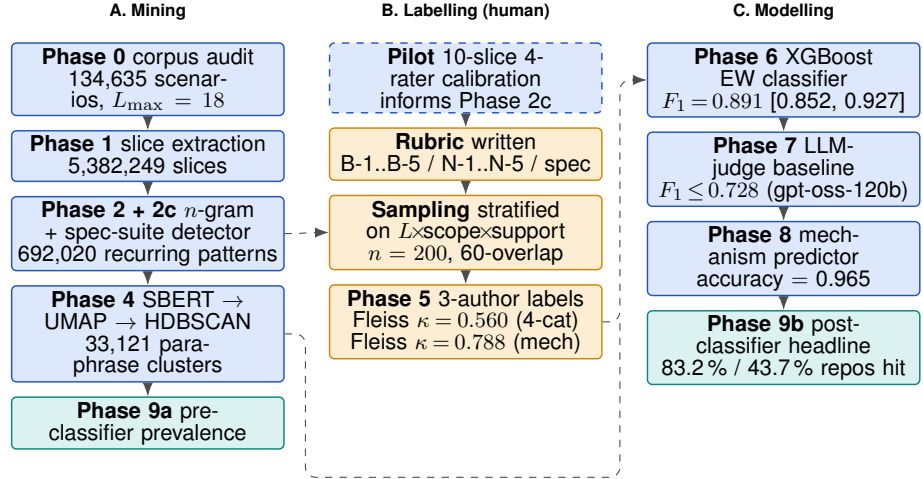

The pipeline is organised into eleven phases, laid out in
Figure~\ref{fig:pipeline}. Phases~0--4 and 9a are fully automated;
Phase~5 (three-author labelling against a written rubric) is the
human critical-path bottleneck; Phases~6, 8, and 9b run after
labels exist.

\subsection{Mining (Phases 0--2c)}
\label{sec:method:phase1}
\label{sec:method:phase2c}

\paragraph{Phase 0: scenario identity.}
Scenarios are canonicalised by the
\texttt{(repo\_slug, file\_path, scenario)} key with
\texttt{is\_background = True} rows excluded, reproducing the
$136{,}970$-scenario count of \citet{mughal2026cukereuse} exactly.
Length percentiles (median 6, p90 12, p95 18, p99 40) fix
$L_{\max}\!=\!18$.

\paragraph{Phase 1: slice extraction.}
For each scenario of step length $S$, every contiguous step window
$(j, j+1, \dots, j+L-1)$ with $L \in [2, \min(S, 18)]$ and
$j \in [0, S{-}L]$ is emitted as a slice carrying
\texttt{(slice\_id, repo\_slug, file\_path, scenario,
position\_start, L, cluster\_id\_seq, text\_seq)}; a 10-step
scenario contributes $\sum_{L=2}^{10}(11{-}L)\!=\!45$ slices and the
corpus contributes 5{,}382{,}249. Steps without an assigned
\texttt{cluster\_id} (7.4\,\% of the corpus) are dropped. Slices are
strictly contiguous ($g\!=\!0$); the gap-tolerant family
\citep{pei2001prefixspan,srikant1996gsp,zaki2001spade,mannila1997episodes}
is left as future work. Any recurrence found \emph{only}
under $g\!\geq\!1$ violates the rubric's stable-context criterion
and is necessarily less extraction-worthy, so the headline
prevalences below are conservative lower bounds.

\paragraph{Phase 2: exact \texorpdfstring{$n$}{n}-gram counting.}
For each unique cluster-id sequence we accumulate
\texttt{support\_total}, \texttt{n\_distinct\_files}, and the per-RQ
scope metrics defined in Section~\ref{sec:approach:scopes}; patterns
with \texttt{support\_total~$<\!2$} are dropped. The output is a
15.3\,MB parquet of 692{,}020 distinct recurring patterns.

\paragraph{Phase 2c: refinement.}
Two pilot-labelling corrections (Section~\ref{sec:discussion}) are
applied: (i)~adding \texttt{n\_distinct\_orgs} (split on the first
underscore of \texttt{repo\_slug}) as the primary RQ3 metric instead
of \texttt{n\_distinct\_repos}, addressing same-owner multi-repo
inflation; (ii)~replacing the v1 spec-suite detector (per-file
pattern density alone) with a v3 detector that requires both high
density and a generator-template signature, namely $>\!50$ distinct
RQ1 patterns AND either top-pattern within-file recurrence $>\!100$
or $\geq\!30\,\%$ of distinct cluster canonical texts containing
two adjacent quoted single-word placeholders. The v3 detector
re-classifies legitimately heavily-duplicated software test code (e.g.,
\texttt{Corvusoft/restq}, \texttt{git-town/git-town}) as real
signal.

\subsection{Phase 4: SBERT slice-embedding clustering}

For each unique cluster-id sequence with a positive RQ scope
signal, a slice embedding is mean-pooled from SBERT
(\texttt{all-MiniLM-L6-v2}, 384-d) embeddings of the canonical
texts of its constituent clusters \citep{reimers2019sbert}, reduced
to 50 dimensions with UMAP \citep{mcinnes2018umap} and clustered
with HDBSCAN \citep{campello2013hdbscan}. On 619{,}827 input
patterns this yields 33{,}121 paraphrase-equivalence clusters
(25.3\,\% noise points; median size 10, p95 35). Hyperparameters
are pinned in the released script.

\subsection{Phase 5: rubric and three-author labelling}
\label{sec:method:phase5}

A 200-slice pool is sampled stratified by $L$-bucket, scope
(most-specific scope wins), and support bucket; 180 slices come from
the real-signal stratum (outlier-fraction $\leq 0.5$) and 20 from a
spec-coverage stratum that probes the \texttt{flagged-spec} edge
case. A 60-slice overlap subset is labelled by all three authors;
the remaining 140 split 46/46/48. The rubric (released under
Apache-2.0\footnote{Rubric and per-author labels at
\href{https://github.com/amughalbscs16/cukereuse_subscenarios_release}{amughalbscs16/cukereuse\_subscenarios\_release};
aggregated labels in
\texttt{methodology/labels.jsonl}.}) defines (a)~a four-category
\emph{extraction-worthy} label (yes / no / uncertain /
flagged-spec) with five positive criteria B-1..B-5 and five
negative criteria N-1..N-5; and (b)~a four-category
\emph{mechanism} label (\texttt{background} /
\texttt{reusable\_scenario} / \texttt{shared\_higher\_level\_step}
/ \texttt{unsure}) conditional on a yes verdict, with \texttt{n/a} recorded otherwise
(these four labels plus \texttt{n/a} are the five categories over which
the mechanism~$\kappa$ is computed). A 10-slice pilot
preceded the main pass and surfaced the three calibration findings
that justified the Phase-2c refinement
(Section~\ref{sec:discussion}).

\subsection{Phases 6--9: classifiers and rollups}

Phase~6 trains an XGBoost binary extraction-worthy classifier on
the labelled pool with bootstrap 95\,\% CIs over the out-of-fold
predictions. Phase~7 re-labels the same pool with two open-weight
LLM judges (Section~\ref{sec:results:llmjudge}). Phase~8 extends
the classifier with a three-way mechanism head, against a
rule-based scope-driven baseline. Phase~9 computes corpus-level
prevalence: 9a is the raw upper bound, and 9b applies the Phase-6
gate to produce the practitioner-facing headline.

\FloatBarrier
\section{Results}
\label{sec:results}

\subsection{Corpus characterisation (Phase 0)}

The cleaned mining set is 134{,}635 named, non-Background,
length-$\geq\!2$ scenarios across 339 repos and 21{,}946
\texttt{.feature} files. Scenarios with fewer than two clustered steps
under the \citet{mughal2026cukereuse} filter cannot match anything and
are dropped (8{,}014 scenarios; 5.9\,\%), leaving 126{,}621 scenarios
as slice input. All 339 repos retain at least one slice. Scenario
length: median 6, p90 12, p95 18, p99 40, max 1{,}373.

\subsection{Slice inventory and recurring patterns (Phases 1, 2 + 2c)}

\begin{figure}[!ht]
\centering
\includegraphics[width=\linewidth]{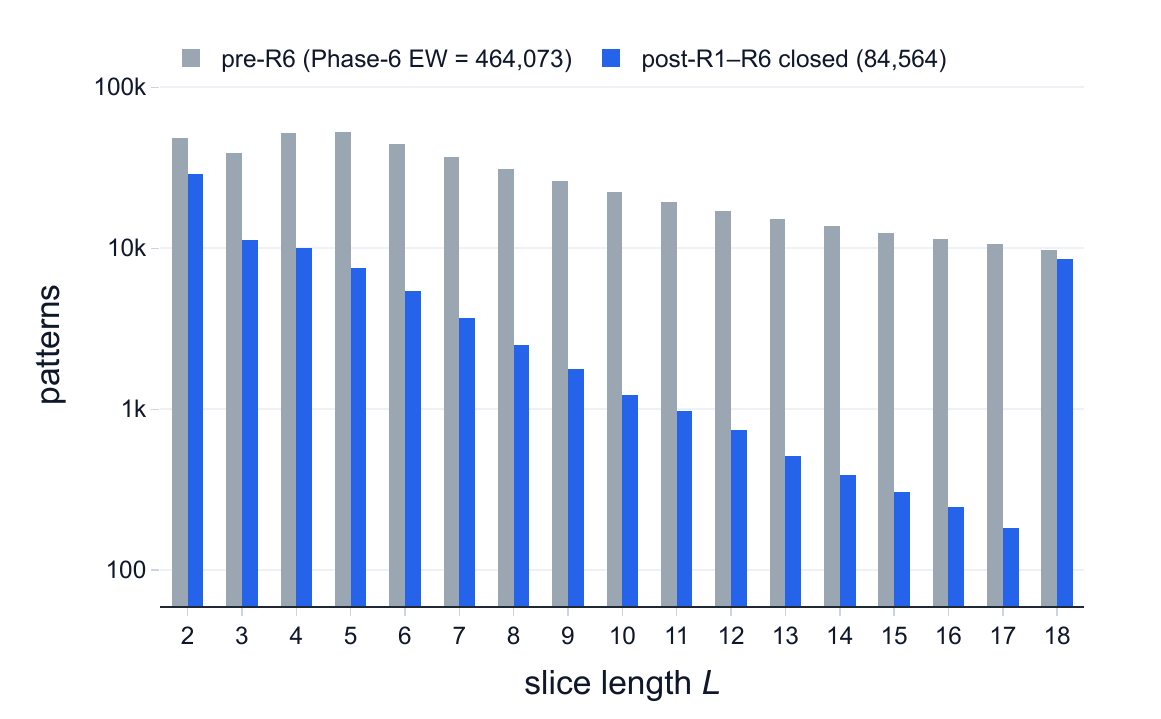}
\caption{Pattern count by slice length $L$ (log-scaled), pre and
post the R1--R6 filter chain.}
\label{fig:Ldist}
\end{figure}

Slice generation yields 5{,}382{,}249 slices, monotonically
decreasing in $L$ ($L=2$: 836{,}737 down through $L=18$: 118{,}069).
After deduplication on cluster-id sequence these collapse to
$2{,}349{,}063$ unique patterns, of which \textbf{692{,}020} have
\texttt{support\_total} $\geq 2$ (the recurring pool).
Figure~\ref{fig:Ldist} shows how this pool distributes across the
17 slice lengths and how the R1--R6 verification filter chain
(Section~\ref{sec:threats}) thins each length-bucket. R6 cannot apply
at $L_{\max}$ (no $L=19$ super-pattern exists), inflating the
post-filter $L=18$ bucket relative to neighbours.

 Within the recurring pool, 403{,}745 patterns carry a
within-file Background signal (\texttt{max\_within\_file\_recurrence}~$\geq 2$),
246{,}007 carry a within-repo reusable-scenario signal
(\texttt{max\_within\_repo\_files}~$\geq 2$), and \textbf{30{,}955
carry a cross-organisational signal}
(\texttt{n\_distinct\_orgs}~$\geq 2$). The cross-owner count
corrects the naive cross-repo metric (62{,}771): 31{,}816 patterns
(51\,\%) of the apparent cross-repo signal are same-owner
cross-repo (e.g., DataDog's multi-language SDK clients).

Table~\ref{tab:top-patterns} shows representative top-ranked patterns
per RQ scope under the real-signal restriction. RQ1 surfaces dense
within-file repetition of background-style invariants; RQ2 surfaces
shared assertion macros within a single repository; RQ3 surfaces
generic HTTP request/response idioms shared across upstream owners.

\begin{table*}[!ht]
\centering
\caption{Representative top-ranked recurring patterns per RQ scope
(real-signal only): RQ1 by
\texttt{max\_within\_file\_recurrence}\,$\times L$; RQ2 by
\texttt{max\_within\_repo\_files}\,$\times L$; RQ3 by
\texttt{n\_distinct\_orgs}.}
\label{tab:top-patterns}
\small
\setlength{\tabcolsep}{4pt}
\begin{tabular}{p{0.04\linewidth}p{0.55\linewidth}p{0.18\linewidth}p{0.13\linewidth}}
\toprule
$L$ & Canonical step text & Signal & Mechanism \\
\midrule
\multicolumn{4}{l}{\emph{RQ1 top: within-file Background candidate}} \\
2 & \texttt{sidekiq should have 0 "event-log" jobs} \newline
    \texttt{sidekiq should have 1 "request-log" job}
  & file-rec=379, support=2{,}504, single repo
  & Background \\
\addlinespace
3 & \texttt{sidekiq should have 1 "webhook" job} \newline
    \texttt{sidekiq should have 1 "event-log" job} \newline
    \texttt{sidekiq should have 1 "request-log" job}
  & file-rec=219, support=854, single repo
  & Background \\
\midrule
\multicolumn{4}{l}{\emph{RQ2 top: within-repo cross-file reusable candidate}} \\
4 & \texttt{the following style:} \newline
    \texttt{the following input:} \newline
    \texttt{I cite all items} \newline
    \texttt{the result should be:}
  & repo-files=438, support=438, 1 org
  & Reusable scenario \\
\addlinespace
4 & \texttt{I generate a type for the schema} \newline
    \texttt{I construct an instance of the schema} \newline
    \texttt{\quad type from the data} \newline
    \texttt{I validate the instance} \newline
    \texttt{the result will be <valid>}
  & repo-files=368, support=1{,}746, 1 org
  & Reusable scenario \\
\midrule
\multicolumn{4}{l}{\emph{RQ3 top: cross-organisational shared candidate}} \\
2 & \texttt{method get} \newline
    \texttt{status 200}
  & 11 orgs, 11 repos, support=4{,}897
  & Shared higher-level step \\
\addlinespace
2 & \texttt{method post} \newline
    \texttt{status 200}
  & 11 orgs, 11 repos, support=3{,}438
  & Shared higher-level step \\
\addlinespace
3 & \texttt{the output should contain:} \newline
    \texttt{the output should not contain:} \newline
    \texttt{the output should not contain:}
  & 8 orgs, 8 repos, support=65
  & Shared higher-level step \\
\bottomrule
\end{tabular}
\end{table*}

\subsection{Slice clustering (Phase 4)}

Of the 619{,}827 patterns with at least one positive scope signal,
HDBSCAN on UMAP-reduced SBERT slice embeddings produces
33{,}121 paraphrase-equivalence clusters. Cluster size is
right-skewed: median 10 patterns, p95 35, max 607. Noise points
account for 25.3\,\% (156{,}617 patterns), i.e., patterns whose
slice embeddings do not cluster densely with any other.

\subsection{Pre-classifier corpus-level prevalence (Phase 9a)}

Two views of scenario-level prevalence: \emph{full} (all recurring
patterns) and \emph{real-signal} (patterns whose majority of
occurrences fall on non-spec-suite files under the Phase-2c v3
filter). The real-signal column is the defensible headline.

\begin{table}[!ht]
\centering
\caption{Corpus-level prevalence by RQ scope across three pruning
stages: \emph{full} (all 692{,}020 recurring patterns),
\emph{real-signal} (non-spec-suite-majority under Phase-2c v3, the
defensible pre-classifier headline), and \emph{post-EW} (after the
Phase-6 extraction-worthy gate). Scenario-level $n$ is over 126{,}621
scenarios; repository-level $n$ is over 339 repos. RQ3 uses
\texttt{n\_distinct\_orgs}~$\geq 2$.}
\label{tab:headline}
\small
\setlength{\tabcolsep}{4pt}
\begin{tabular}{lrrrrrr}
\toprule
& \multicolumn{2}{c}{Full} & \multicolumn{2}{c}{Real-signal} & \multicolumn{2}{c}{Post-EW} \\
\cmidrule(lr){2-3}\cmidrule(lr){4-5}\cmidrule(lr){6-7}
Scope & $n$ & \% & $n$ & \% & $n$ & \% \\
\midrule
Scenarios w/ $\geq 1$ recurring slice & 121{,}701 & 96.1 & --- & --- & --- & --- \\
RQ1 (within-file Background)        & 114{,}131 & 90.1 & 95{,}126 & 75.1 & 95{,}007 & 75.0 \\
RQ2 (within-repo reusable)          & 106{,}496 & 84.1 & 87{,}585 & 69.2 & 75{,}396 & 59.5 \\
RQ3 (cross-org shared-step)         &  21{,}690 & 17.1 & 21{,}690 & 17.1 & 14{,}864 & 11.7 \\
\midrule
\multicolumn{7}{l}{\emph{Repository-level (n=339 repos):}} \\
Repos w/ $\geq 1$ RQ2 candidate     &       300 & 88.5 & --- & --- &      282 & 83.2 \\
Repos w/ $\geq 1$ RQ3 candidate     &       167 & 49.3 & --- & --- &      148 & 43.7 \\
\bottomrule
\end{tabular}
\end{table}

Recurring structure is pervasive within-file and within-repo
($\sim$75\,\% and $\sim$70\,\% of scenarios respectively under the
real-signal restriction); cross-organisational recurrence is rarer
($\sim$17\,\%) but non-trivial. The full-vs-real-signal gap is
largest at RQ1 (90.1\,\%~$\to$~75.1\,\%) and negligible at RQ3
(17.1\,\% in both), confirming that spec-suite generation creates
dense within-file recurrence but rarely escapes its originating
upstream owner.

The v3 spec-suite detector (Section~\ref{sec:method:phase2c}) shrinks
the outlier list from 2{,}459 (11.2\,\%) to 154 files (0.7\,\%),
dominated by \texttt{local-web-services/\allowbreak local-web-services}
(121 files) and the DataDog application-programming-interface
(API) client suites; reclassifying the
2{,}305 dropped files raises the real-signal pattern count from
179{,}019 (v1) to 616{,}464 (v3) of the 692{,}020 recurring patterns.

\subsection{Labelling results (Phase 5)}
\label{sec:results:labels}

Three authors labelled the 200-slice pool under the
Section~\ref{sec:method:phase5} rubric; aggregated distribution and
inter-rater agreement are in Table~\ref{tab:labels}.

\begin{table}[!ht]
\centering
\caption{Three-author labelling outcomes on the 200-slice stratified
pool (60-slice overlap, 140 split 46/46/48). Fleiss'
$\kappa$ \citep{fleiss1971} is computed over the overlap under the
four-category extraction-worthy and five-category mechanism labels
(\texttt{n/a} included for non-yes verdicts).}
\label{tab:labels}
\small
\begin{tabular}{lrrr}
\toprule
& Author A & Author B & Author C \\
\midrule
\multicolumn{4}{l}{\emph{Per-author extraction-worthy distribution:}} \\
\quad yes / no / uncertain / flagged-spec & 77/12/4/13 & 72/11/11/12 & 72/23/0/13 \\
\quad total                              & 106 & 106 & 108 \\
\midrule
\multicolumn{4}{l}{\emph{Inter-rater agreement on the 60-slice overlap:}} \\
\quad Fleiss' $\kappa$ (extract 4-cat / mech 5-cat) & \multicolumn{3}{r}{0.560 / 0.788} \\
\quad Pairwise A--B / A--C / B--C (extract)         & 0.717 & 0.850 & 0.750 \\
\quad Pairwise A--B / A--C / B--C (mech)            & 0.800 & 0.933 & 0.800 \\
\midrule
\multicolumn{4}{l}{\emph{Majority verdict on the overlap (extraction-worthy):}} \\
\quad yes / no / flagged-spec / 3-way tie & \multicolumn{3}{r}{41 / 9 / 7 / 3} \\
\bottomrule
\end{tabular}
\end{table}

Under the Landis--Koch interpretation
\citep{landis1977kappa}, the four-category extraction-worthy
$\kappa = 0.56$ is \emph{moderate} and the five-category mechanism
$\kappa = 0.79$ is \emph{substantial}. Most extraction-worthy
disagreement concentrates on three failure modes: (a)~the
yes-vs-no boundary on $L\!=\!2$ cross-organisational trivial
content (e.g., \texttt{method post}/\texttt{status 200}), where
the rubric's worked Example~3 explicitly admits a borderline
call; (b)~the \texttt{uncertain} bucket, used 0--11 times by
different authors; and (c)~the \texttt{flagged-spec} vs.\
\texttt{no} boundary on placeholder-free heavily-duplicated
fixtures. Three slices ended in three-way ties (released as
\texttt{tie}, not tie-broken). The mechanism label follows
almost mechanically from scope once extraction-worthiness is
established, explaining its higher $\kappa$.

Excluding the spec-coverage stratum, the overlap-majority breakdown
(Table~\ref{tab:labels}, last row) is consistent with the population that
survives the Phase-2c v3 filter: most surviving patterns are
extraction-worthy, and the rest concentrate in the
trivial-content tail at low $L$.

\subsection{Extraction-worthy classifier (Phase 6)}
\label{sec:results:classifier}

An XGBoost binary classifier \citep{chen2016xgboost}
(\texttt{n\_estimators=200}, \texttt{max\_depth=4},
\texttt{learning\_rate=0.1}) is fit on the 197 non-tie labelled slices.
Positive class = \texttt{yes}; negative = \texttt{no}~$\cup$~\texttt{uncertain}~$\cup$~\texttt{flagged-spec}.
Features: $L$, \texttt{support\_total}, the three count features
(\texttt{n\_distinct\_repos/orgs/files}),
\texttt{max\_within\_file\_recurrence}, \texttt{max\_within\_repo\_files},
\texttt{outlier\_fraction}, \texttt{has\_template\_structure},
scope one-hot flags (e.g.\ \texttt{scope\_RQ2}), and three derived
ratios including \texttt{ratio\_within\_repo} and
\texttt{ratio\_within\_file}. Evaluation is 5-fold
stratified cross-validation (CV) with $1{,}000$-bootstrap 95\,\%
percentile confidence intervals (CIs)
\citep{efron1993bootstrap}. The trained classifier is applied to the
\emph{scope-eligible} pattern population: those satisfying at least
one of $\{$RQ1, RQ2, RQ3-cross-org$\}$ ($n = 595{,}857$), a strict
subset of the $619{,}827$ patterns clustered in Phase~4. The
$23{,}970$-pattern residual is same-owner cross-repo recurrence
(RQ3 by the naive cross-repo metric but not by
\texttt{n\_distinct\_orgs}~$\geq 2$), which maps to no Mughal-2024
mechanism and is excluded from classifier input.

\begin{table}[!ht]
\centering
\caption{Phase-6 extraction-worthy classifier (XGBoost, binary).
Out-of-fold metrics from 5-fold stratified CV over 197 non-tie labels;
95\,\% CIs are $1{,}000$-bootstrap percentile intervals.}
\label{tab:classifier}
\small
\begin{tabular}{lrr}
\toprule
Metric & Median & 95\,\% CI \\
\midrule
Precision & 0.868 & [0.811, 0.920] \\
Recall    & 0.916 & [0.869, 0.959] \\
F$_1$     & 0.891 & [0.852, 0.927] \\
ROC-AUC   & 0.881 & [0.818, 0.931] \\
\bottomrule
\end{tabular}
\end{table}

\begin{figure}[!ht]
\centering
\includegraphics[width=\linewidth]{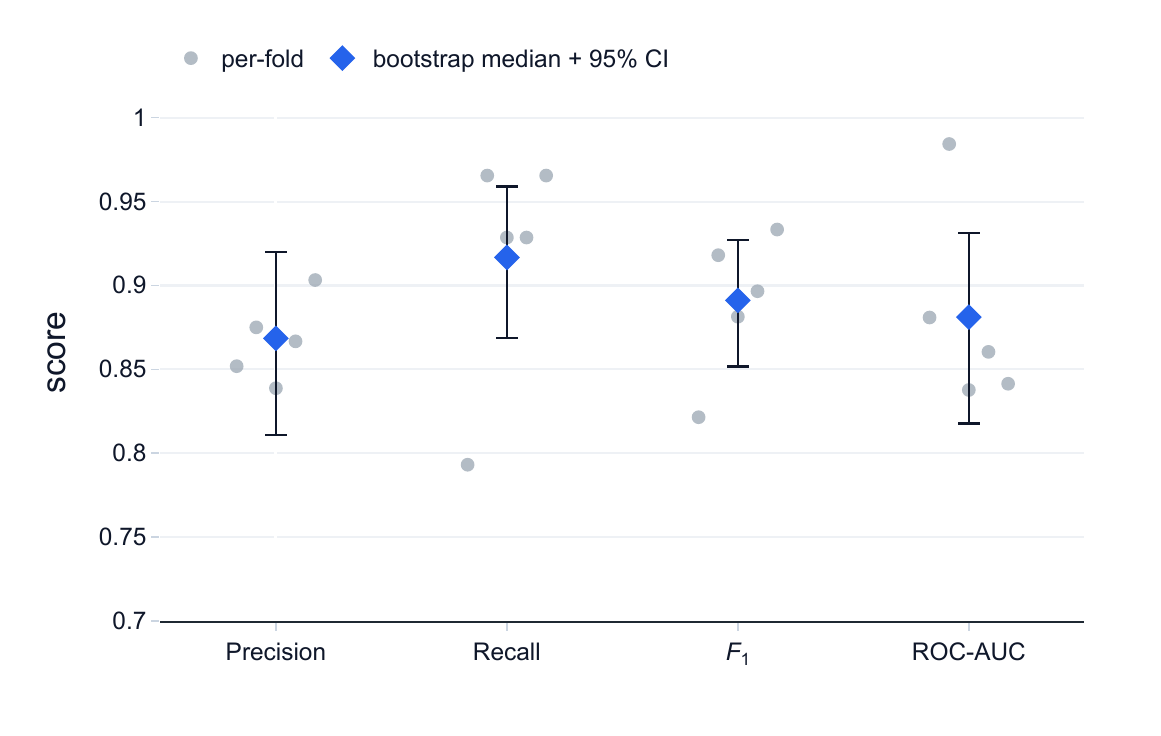}
\caption{Phase-6 classifier: per-fold precision, recall, $F_1$, and
ROC-AUC across 5-fold stratified CV, overlaid with the bootstrap
median and 95\,\% CI from Table~\ref{tab:classifier}.}
\label{fig:fold-metrics}
\end{figure}

The classifier reaches an out-of-fold $F_1 = 0.891$
(95\,\% CI [0.852, 0.927]) and bootstrap-median area under the
receiver-operating-characteristic curve (ROC-AUC) $= 0.881$.
Figure~\ref{fig:fold-metrics} plots per-fold scores against the
bootstrap median and 95\,\% CI; the across-fold variance is
modest, with fold-1 the lower outlier on F$_1$ and ROC-AUC.

\begin{figure}[!ht]
\centering
\includegraphics[width=\linewidth]{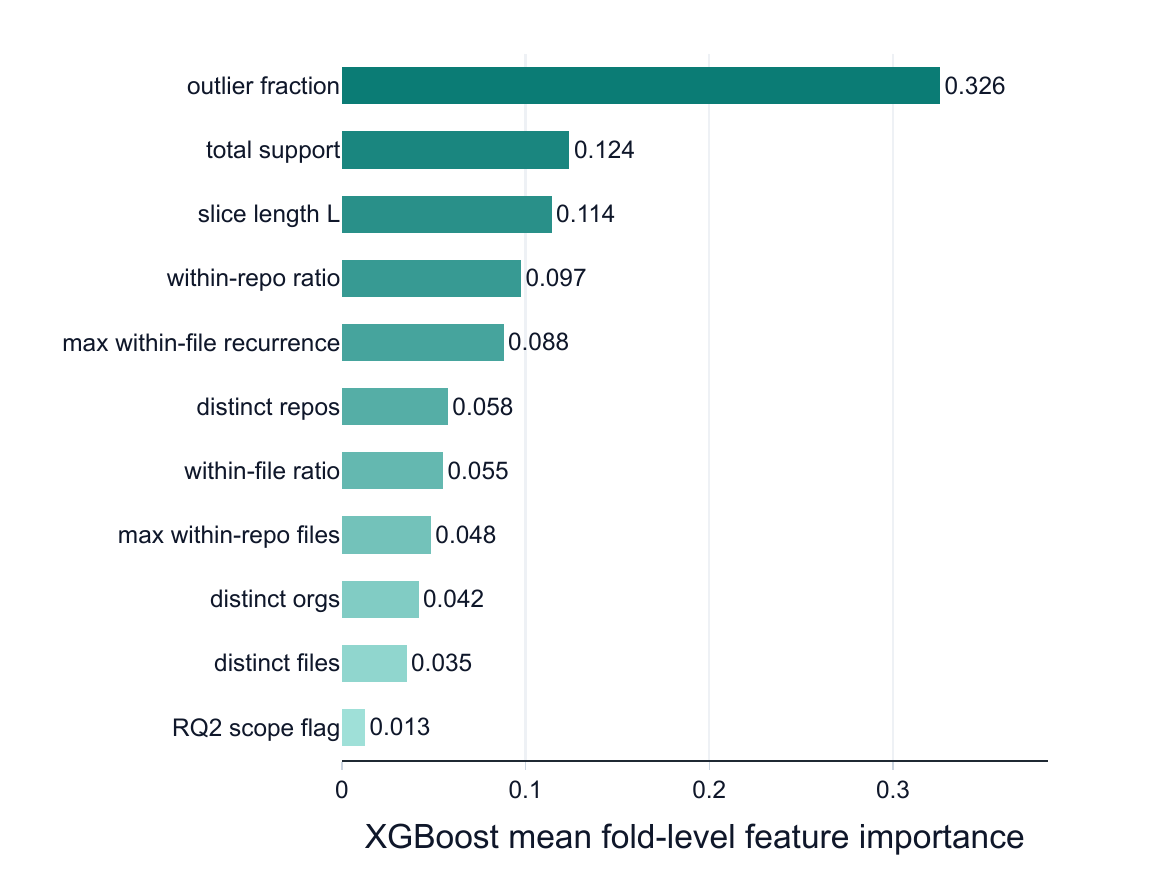}
\caption{Phase-6 classifier feature importance, mean across the
five XGBoost folds. Three feature families dominate: spec-suite signal
(\texttt{outlier\_fraction}), recurrence density
(\texttt{support\_total}, $L$, \texttt{max\_within\_file\_recurrence}),
and structural ratios.}
\label{fig:feature-importance}
\end{figure}

\paragraph{Non-ML rule baselines.}
To test whether XGBoost earns its added complexity, it is compared
against two baselines on the same 197-item out-of-fold setting. The trivial
\emph{all-yes} predictor reaches $F_1 = 0.841$ (driven by the 72.6\,\%
base rate of \texttt{yes} on the labelled pool); a single-feature
rule \texttt{outlier\_fraction}~$<\!0.3$ reaches $F_1 = 0.836$
($P=0.801$, $R=0.874$). XGBoost's lift over the rule is concentrated
in precision ($P=0.868$ vs $0.801$), with recall roughly matched. On
the 45 discordant items (31 XGBoost-right, 14 rule-right) McNemar's
test gives $\chi^2 = 5.69$, $p = 0.017$: the lift is statistically
significant though modest. Because the pool is imbalanced, yes-class
$F_1$ understates what the gate contributes. On the same out-of-fold
predictions XGBoost attains a Matthews correlation coefficient (MCC)
of $0.575$, balanced accuracy $0.773$, and negative-class precision /
recall of $0.739 / 0.630$. The all-yes predictor scores
MCC $= 0$, balanced accuracy $0.500$, and negative-class recall $0$
by construction: it can never reject a candidate, which is the
gate's entire purpose. McNemar against all-yes on the 46 discordant
items (34 XGBoost-right, 12 all-yes-right) gives $\chi^2 = 9.59$,
$p = 0.002$.
Feature importance (Figure~\ref{fig:feature-importance}) is led by
\texttt{outlier\_fraction}, whose dominance reflects the Phase-2c
spec-suite signal, with density features (\texttt{support\_total},
$L$, \texttt{max\_within\_file\_recurrence}) as a secondary cluster.
Scope one-hots are near-zero because $L$ and the density features
already encode scope. After per-fold
evaluation, we re-fit on all 197 non-tie labels and apply the model
to the 595{,}857 scope-eligible patterns, of which 464{,}073
(77.9\,\%) are predicted extraction-worthy.

\subsection{Mechanism predictor (Phase 8)}
\label{sec:results:mechanism}

Conditional on a slice being labelled extraction-worthy, the
mechanism label predicts which of the three Mughal-2024 targets
(\texttt{background}, \texttt{reusable\_scenario},
\texttt{shared\_higher\_\allowbreak level\_step}) is appropriate.
Among the 143 \texttt{yes}-labelled slices, the mechanism distribution
is 24 \texttt{background}, 70 \texttt{reusable\_scenario}, and 49
\texttt{shared\_higher\_\allowbreak level\_step}; four \texttt{unsure}
verdicts on \texttt{yes}-labelled slices were resolved by majority
vote during aggregation.

\paragraph{Rule-based scope-driven baseline.}
The natural rule-based predictor maps RQ1 $\to$ \texttt{background},
RQ2 $\to$ \texttt{reusable\_scenario}, and RQ3 $\to$
\texttt{shared\_higher\_level\_step}. On the 143 labels this baseline
achieves accuracy~$0.972$ and macro-F$_1 = 0.965$, with the four
mismatches all in the RQ1 bucket where the labellers preferred
\texttt{reusable\_scenario} over \texttt{background} (a slice that
recurred in one file plus one other file qualifies for both
mechanisms; the labellers preferred the cross-file mechanism for
those four cases).

\paragraph{Learned multi-class XGBoost predictor.}
A multi-class XGBoost classifier
(same features as Phase~6, \texttt{objective=multi:softprob}) reaches
out-of-fold accuracy $0.965$ and macro-F$_1 = 0.955$ in 5-fold
stratified cross-validation, statistically indistinguishable from
the rule-based baseline at this sample size. Per-class precision /
recall on out-of-fold predictions are $0.85 / 0.96$ for
\texttt{background}, $0.99 / 0.94$ for \texttt{reusable\_scenario},
and $1.00 / 1.00$ for \texttt{shared\_higher\_level\_step}. On the
labelled out-of-fold sample the
\texttt{shared\_higher\_level\_step} class is perfectly separable
because cross-organisational recurrence
($\texttt{n\_distinct\_orgs} \geq 2$) is a precondition for the class
under the rubric; the learned model does not enforce that constraint
at deployment, where filter R5 (Section~\ref{sec:threats}) removes
the $5{,}945$ predictions that violate it.

\begin{figure}[!ht]
\centering
\includegraphics[width=\linewidth]{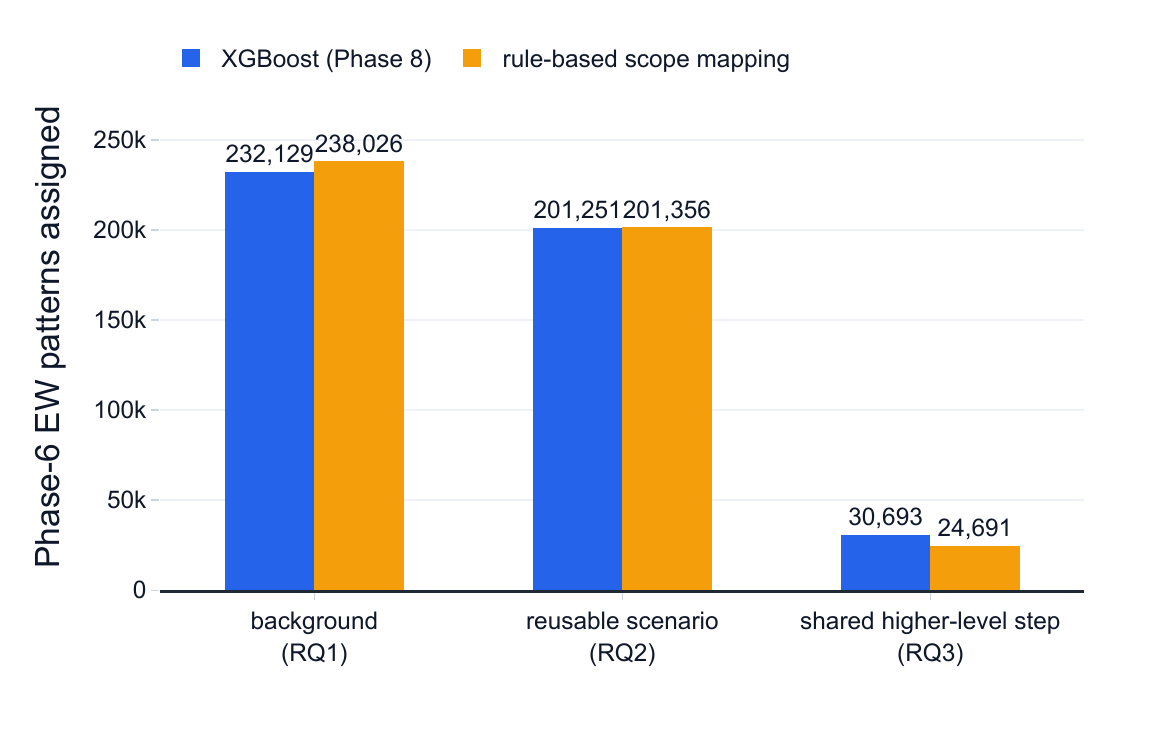}
\caption{Phase-8 mechanism distribution over the 464{,}073 predicted-EW
patterns: XGBoost vs the rule-based scope-mapping baseline.}
\label{fig:mech}
\end{figure}

\paragraph{Application to the 464{,}073 predicted-extraction-worthy
patterns.} Re-fitting on all 143 labels and applying to the
predicted-EW population yields 232{,}129
\texttt{background} (50.0\,\%), 201{,}251
\texttt{reusable\_scenario} (43.4\,\%), and 30{,}693
\texttt{shared\_higher\_level\_step} (6.6\,\%). XGBoost and the
rule-based baseline agree on $98.6\,\%$ of patterns; differences
concentrate at the
\texttt{background}~$\leftrightarrow$~\texttt{shared\_higher\_level\_step}
boundary, where the learned predictor exploits a secondary cross-org
signal. Both predictors are released so a downstream
command-line tool (CLI) can pick its preferred one. Figure~\ref{fig:mech} shows the full mechanism
distribution for both predictors side by side.

\subsection{LLM-judge baseline (Phase 7)}
\label{sec:results:llmjudge}

To pre-empt the \emph{did-you-try-an-LLM-baseline?} reviewer concern,
the 200-slice pool is re-labelled with two open-weight LLMs via
OpenRouter \citep{zheng2023llmjudge,gu2026llmjudgesurvey} at
\texttt{temperature = 0}: \texttt{openai/gpt-oss-120b} (120\,B) and
\texttt{inclusionai/ling-2.6-1t} (1\,T-parameter MoE). Each query
supplies the condensed rubric (B-1..B-5, N-1..N-5, spec-suite handling,
calibration notes), the slice's $L$-step canonical text, the per-scope
recurrence signals, and \texttt{outlier\_fraction} from Phase~2c. Both
models return a JSON verdict with \texttt{extraction\_worthy} and
\texttt{mechanism} fields; the parser tolerates markdown fences and
prose around the JSON.

\paragraph{Agreement results.}
\texttt{gpt-oss-120b} reaches $\kappa = 0.348$ (fair) with
$F_1(\text{yes}) = 0.728$; \texttt{ling-2.6-1t} reaches
$\kappa = 0.243$ with $F_1(\text{yes}) = 0.587$
(Table~\ref{tab:llmjudge}). Both models are highly precise on the yes
class ($P = 0.91$ / $0.94$) but conservative in recall ($R = 0.61$ /
$0.43$), under-calling extraction-worthiness and over-calling
\emph{no} or \emph{flagged-spec}. Inter-LLM Fleiss'~$\kappa$ on the full
200 patterns is $0.393$ (4-cat) / $0.304$ (binary), materially below
the human triad's $0.560$ (4-cat) and pairwise raw agreement
$0.717$--$0.850$ (Table~\ref{tab:labels}). On the conditional
mechanism-given-yes task both models reach $\geq 0.95$ accuracy: once
the extraction-worthy gate is passed, the three-way mechanism call is
the easier sub-task.

\begin{table}[!ht]
\centering
\caption{LLM-judge baseline (Phase 7) on the two full-coverage
open-weight models. $n_v$ counts parseable verdicts. Binary metrics
(acc$_b$, $\kappa_b$, $F_1$(yes)) are versus the human aggregated label
on the $n{=}197$ non-tie subset; mech.\ is mechanism accuracy
conditional on both rater and model saying \emph{yes}.}
\label{tab:llmjudge}
\small
\begin{tabular}{lrrrrr}
\toprule
Rater                 & $n_v$ & acc$_b$ & $\kappa_b$ & $F_1$(yes) & mech. \\
\midrule
\texttt{gpt-oss-120b} & 200 & 0.67 & 0.348 & 0.728 & 0.977 \\
\texttt{ling-2.6-1t}  & 200 & 0.56 & 0.243 & 0.587 & 0.951 \\
\bottomrule
\end{tabular}
\end{table}

\begin{figure}[!ht]
\centering
\includegraphics[width=\linewidth]{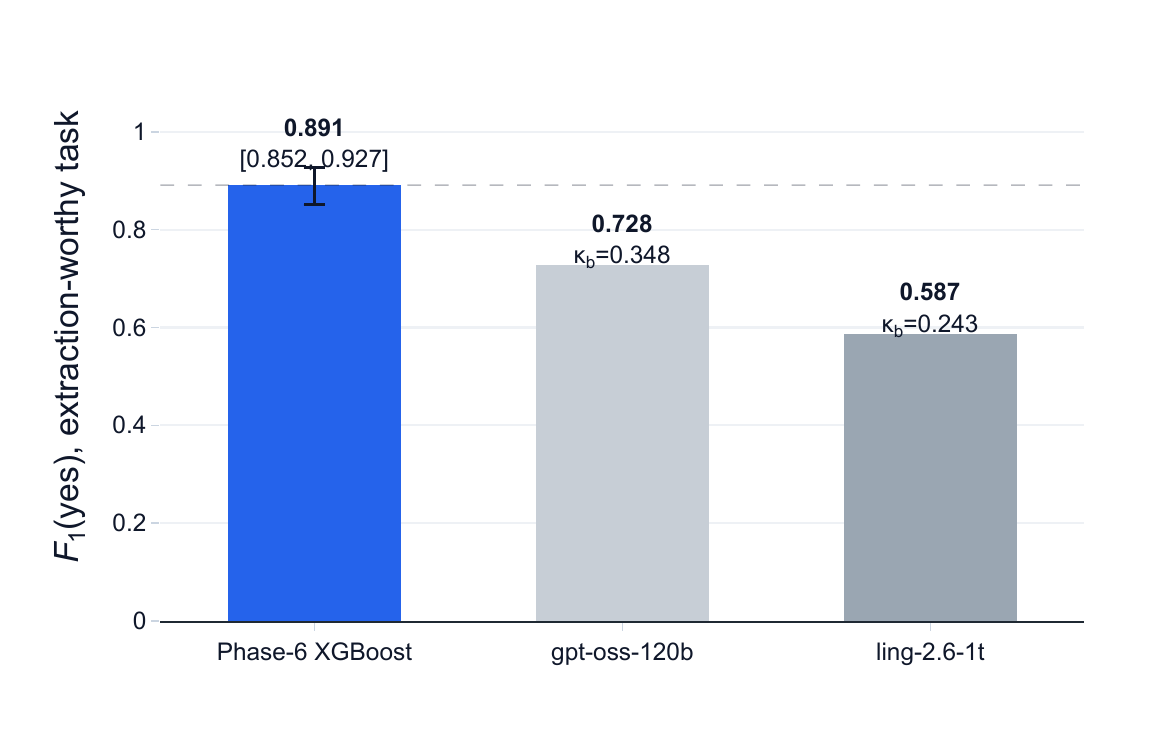}
\caption{Yes-class $F_1$ on the binary extraction-worthy task vs the
human aggregated label (197 non-tie items). Phase-6 classifier reaches
$F_1 = 0.891$; the better LLM judge reaches $F_1 = 0.728$. Error bar
is the classifier bootstrap 95\,\% CI; LLM annotations show Cohen's
$\kappa_b$.}
\label{fig:cls-vs-llm}
\end{figure}

\paragraph{Implication for deployment.}
\emph{LLM-as-judge with off-the-shelf open-weight models does not
match the Phase-6 classifier} on this task: $F_1 = 0.891$ (95\,\% CI
$[0.852, 0.927]$) against $F_1 = 0.728$ / $\kappa_b = 0.348$ for the
better LLM. McNemar's test on the discordant out-of-fold predictions
confirms the gap is statistically significant: against
\texttt{gpt-oss-120b}, XGBoost is right-only on 52 items and the LLM
is right-only on 19 ($\chi^2 = 14.4$, $p = 1.5\!\times\!10^{-4}$);
against \texttt{ling-2.6-1t}, 71 vs 17 ($\chi^2 = 31.9$,
$p < 10^{-4}$). The classifier costs one 200-slice three-author label
pool plus a CPU-minute to fit; scoring the $595{,}857$ scope-eligible
patterns is a single batch \texttt{predict\_proba} call. The Phase-6
classifier therefore remains the primary gate; the LLM-judge numbers
stand as a methodological reference (Figure~\ref{fig:cls-vs-llm}).

\subsection{Post-classifier corpus headline (Phase 9b)}
\label{sec:results:phase9b}

The post-EW columns of Table~\ref{tab:headline} report the same
prevalences after the Phase-6 gate. The classifier prunes 0.1\,pp from
RQ1, 9.7\,pp from RQ2, and 5.4\,pp from RQ3; pruning concentrates at
RQ2 and RQ3 because
most filtered candidates are L=2 trivial-content cross-org HTTP pairs
or within-repo low-content macros that the rubric flags under N-4
(slice-too-short-for-value). At repository scale, 83.2\,\% of repos
still host an EW reusable-scenario candidate and 43.7\,\% host an EW
cross-org shared-step candidate.

\paragraph{Threshold sensitivity and uncertainty.}
The post-EW figures are model-based point estimates at the default
$0.5$ probability cutoff and inherit the Phase-6 error rates; they
should be read as bracketed by the full and real-signal columns of
Table~\ref{tab:headline}. Sweeping the decision threshold from $0.3$
to $0.7$ keeps scenario-level prevalence within
$67.1$--$77.6\,\%$ (RQ1), $55.5$--$63.0\,\%$ (RQ2), and
$10.3$--$14.0\,\%$ (RQ3), and repository-level prevalence within
$82.9$--$84.1\,\%$ (RQ2) and $41.3$--$44.5\,\%$ (RQ3), so the
qualitative conclusions do not depend on the cutoff. The sweep
script and per-threshold rollups are released with the artefacts.

\FloatBarrier
\section{Discussion}
\label{sec:discussion}

\subsection{Pilot labelling and rubric calibration}

A 10-slice pilot surfaced three calibration findings.
\textbf{Finding~1 (owner vs repo).} Two RQ3 entries with
\texttt{n\_distinct\_repos}~$=5$ were one upstream owner's multi-language
SDK clients (e.g., DataDog's Go/Java/Python/Ruby/TS): cross-repo but
not cross-owner. \texttt{n\_distinct\_orgs} was therefore adopted as
the primary RQ3 metric.
\textbf{Finding~2 (long-$L$ sub-extraction).} A length-18 pilot slice
(Kolibri coach lesson-report workflow) contained four repetitions of
an inner 4-step pattern; the right target is the inner pattern,
not the enclosing block. The current rubric is binary on the slice as
given; sub-slice preference is future work.
\textbf{Finding~3 (spec-suite v1 over-broad).}
A heavily-duplicated but template-free pilot slice
(\texttt{Corvusoft/restq}, support 786) was wrongly flagged by the v1
detector. The v3 detector (Section~\ref{sec:method:phase2c}) requires
both density and a template-structure signature, sharply shrinking the
outlier list.
All three findings translate directly into Phase-6 classifier
features: a template-structure flag (Finding~3) and
\texttt{n\_distinct\_orgs} distinct from \texttt{n\_distinct\_repos}
(Finding~1) are implemented; sub-slice preference (Finding~2)
remains future work (Phase~2.5).

\subsection{Cross-organisational signal magnitude}

\begin{figure}[!ht]
\centering
\includegraphics[width=\linewidth]{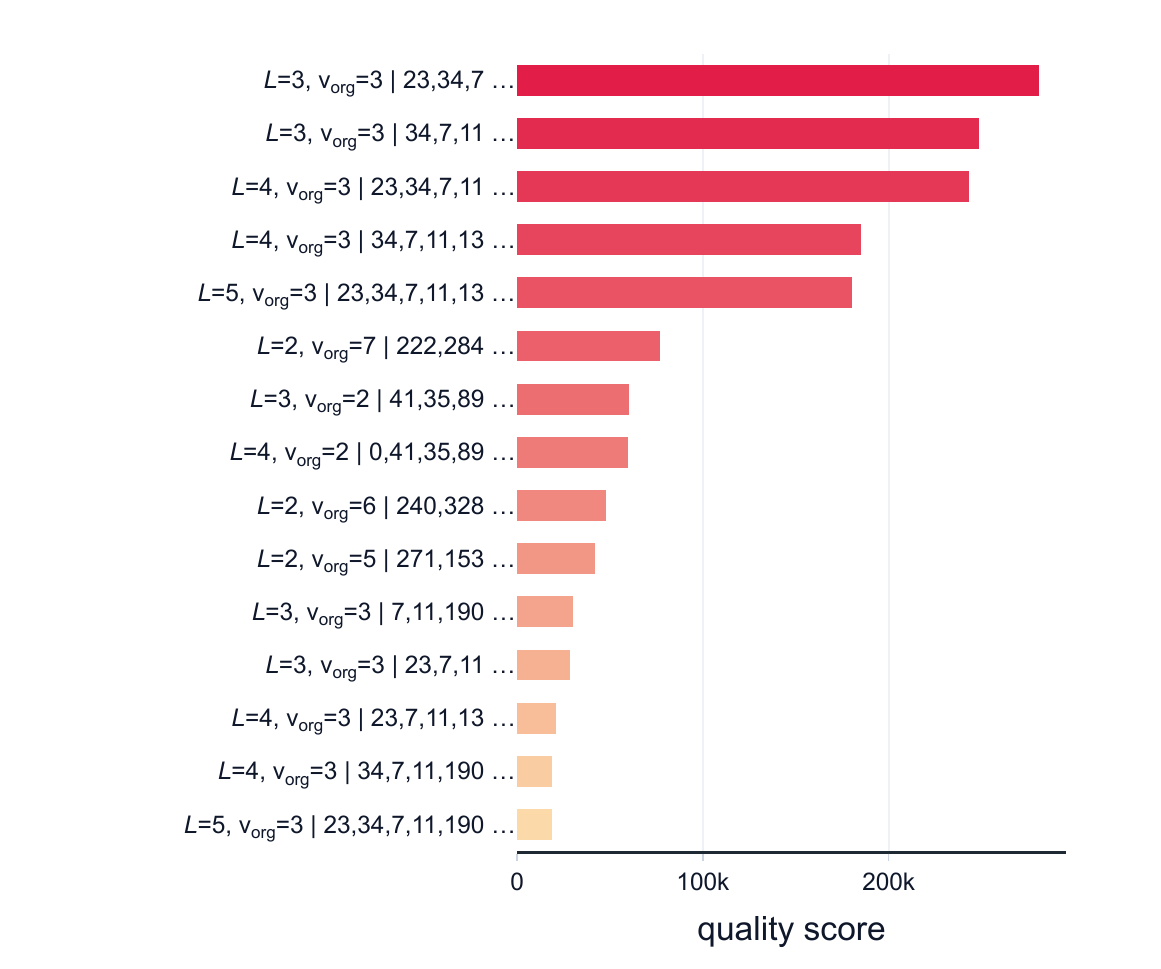}
\caption{Top 20 cross-organisational (RQ3) candidates by quality
score ($n_{\mathrm{scen}} \cdot \sqrt{\mathrm{support}} \cdot
n_{\mathrm{orgs}}$, the ranking key released with the candidates)
after the R6 closure filter (Section~\ref{sec:threats}).
Fingerprints are abbreviated to the first six
cluster ids.}
\label{fig:top-rq3}
\end{figure}

The 30{,}955 RQ3 candidates are 4.5\,\% of the recurring pool but
carry disproportionate practical interest: they are the only
candidates for which the shared-higher-level-step mechanism is the
appropriate target. The leaderboard (Figure~\ref{fig:top-rq3}) is
dominated by what \citet{binamungu2018saner} call
\emph{infrastructural} duplication (HTTP request-response idioms
such as \texttt{method get / status 200} across 11 distinct upstream
owners, and CLI output assertions) rather than domain duplication. The returns drop sharply past $L=3$ (Figure~\ref{fig:top-rq3}), so the
shared-higher-level-step mechanism is most useful for short,
frequently-repeated infrastructural macros rather than long
business-logic sequences.

\subsection{Industrial relevance}

\citet{irshad2022ist} report that BDD software-test specification refactoring is
a recurring but under-tooled industrial task in which the manual cost
of identifying targets is the limiting factor; \citet{kim2014refactoring}
find the same cost asymmetry at Microsoft on production code, where
discovery and prioritisation dominate. The post-EW columns of
Table~\ref{tab:headline} are the practitioner-facing prior: the median
repo contains 115 recurring patterns (p25 = 23, p75 = 555) before the
gate. A
practitioner running \subtool on their own repository can therefore
expect a non-empty report of EW-classified candidates with high
probability (83.2\,\% of corpus repos host at least one EW
reusable-scenario candidate; Table~\ref{tab:headline}).

\paragraph{Inspection-burden framing.}
The scale argument is hypothetical rather than measured: if a
reviewer were to budget a nominal 30~s per candidate (a figure we
assume rather than measure), exhaustive triage of the $464{,}073$
predicted-EW patterns would be infeasible ($\sim$3{,}900 reviewer-hours),
and even the $84{,}564$ R1--R6 survivors
(Section~\ref{sec:threats}) would require $\sim$700 hours.
Ranking-by-quality-score therefore makes a fixed top-$K$ review
budget possible ($K{=}200 \approx 1.7$ hours), shrinking the
inspection \emph{set} by $\sim$2{,}300$\times$ relative to the
unfiltered population ($\sim$420$\times$ relative to the R1--R6
survivors). Whether the top-$K$ set captures the candidates a
maintainer would actually act on (recall at $K$) is unmeasured; it
is the subject of the field study in future work~(ii). The
per-candidate ranking is released so any $K$ can be chosen to match
the reviewer budget.

\FloatBarrier
\section{Threats to validity}
\label{sec:threats}
\label{sec:limitations}

\subsection{Internal validity}

\paragraph{Cluster-id sequence collisions.}
Two semantically distinct slices may share a cluster-id sequence if the
\cukereuse hybrid clusterer over-merges their steps.
\citet{mughal2026cukereuse} report Fleiss' $\kappa = 0.84$ on a
1{,}020-pair benchmark, which bounds the within-cluster confusion rate
but not the slice-level collision rate (a slice is correct only if
every constituent step is correctly clustered). The mitigation is
manual inspection of the top-100 ranked slices per RQ scope; the audit
log is released alongside the ranking parquet. Phase~4 catches the
inverse direction (semantically equivalent slices with divergent
cluster-id sequences) by collapsing them into a paraphrase-equivalence
cluster.

\paragraph{Subjectivity of \emph{extraction-worthy}.}
The Phase~6 classifier learns from human judgements that may not
generalise. Fleiss' $\kappa$ \citep{fleiss1971} on the 60-slice overlap
subset is the inter-rater agreement floor, interpreted under the same
Landis--Koch bands \citep{landis1977kappa} as \cukereuse; the pair-level
rubric achieved $\kappa = 0.84$, the calibration target for the
slice-level extension.

\paragraph{Detector-threshold sensitivity.}
The Phase-2c v3 spec-suite detector requires file-level density
$> 50$ AND either top-pattern within-file recurrence $> 100$ OR
template-structure fraction $\geq 0.30$. These thresholds were
calibrated on three pilot entries (Section~\ref{sec:discussion}) and
on visual inspection of the file-level density histogram.
Table~\ref{tab:headline} reports the headline both with and without
the v3 filter so a reader who disagrees with the threshold choice can
recover a defensible interval; a $\pm 50\,\%$ sensitivity sweep is in
the supplementary material. An analogous sweep of the Phase-6
classifier decision threshold is reported in
Section~\ref{sec:results:phase9b}.

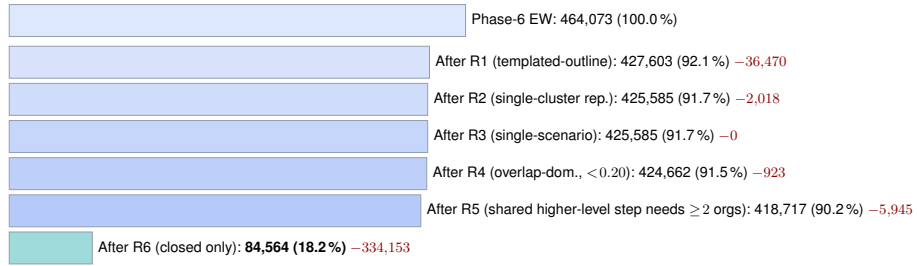
\begin{figure}[!ht]
\centering
\resizebox{0.92\linewidth}{!}{\sffamily%
\begin{tikzpicture}[
  every node/.style={font=\footnotesize},
  band/.style={rectangle, draw=sngrey!70, line width=0.4pt,
               minimum height=7mm, anchor=west},
  drop/.style={font=\scriptsize\itshape, text=red!60!black},
  label/.style={font=\scriptsize, anchor=west},
]
\node[band, fill=snblue!15, minimum width=100mm] (b0) at (0,0) {};
\node[label,anchor=west] at (b0.east) {Phase-6 EW: 464{,}073 (100.0\,\%)};

\node[band, fill=snblue!18, minimum width=92.1mm, below=2mm of b0.south west, anchor=north west] (b1) {};
\node[label] at (b1.east) {After R1 (templated-outline): 427{,}603 (92.1\,\%) \textcolor{red!55!black}{\scriptsize $-36{,}470$}};

\node[band, fill=snblue!22, minimum width=91.7mm, below=1mm of b1.south west, anchor=north west] (b2) {};
\node[label] at (b2.east) {After R2 (single-cluster rep.): 425{,}585 (91.7\,\%) \textcolor{red!55!black}{\scriptsize $-2{,}018$}};

\node[band, fill=snblue!26, minimum width=91.7mm, below=1mm of b2.south west, anchor=north west] (b3) {};
\node[label] at (b3.east) {After R3 (single-scenario): 425{,}585 (91.7\,\%) \textcolor{red!55!black}{\scriptsize $-0$}};

\node[band, fill=snblue!30, minimum width=91.5mm, below=1mm of b3.south west, anchor=north west] (b4) {};
\node[label] at (b4.east) {After R4 (overlap-dom., $<\!0.20$): 424{,}662 (91.5\,\%) \textcolor{red!55!black}{\scriptsize $-923$}};

\node[band, fill=snblue!34, minimum width=90.2mm, below=1mm of b4.south west, anchor=north west] (b5) {};
\node[label] at (b5.east) {After R5 (shared higher-level step needs $\geq\!2$ orgs): 418{,}717 (90.2\,\%) \textcolor{red!55!black}{\scriptsize $-5{,}945$}};

\node[band, fill=snteal!40, minimum width=18.2mm, below=1mm of b5.south west, anchor=north west] (b6) {};
\node[label] at (b6.east) {After R6 (closed only): \textbf{84{,}564 (18.2\,\%)} \textcolor{red!55!black}{\scriptsize $-334{,}153$}};
\end{tikzpicture}}
\caption{Six-rule verification filter funnel on the $464{,}073$
Phase-6 EW candidates (bar width $\propto$ surviving count). Per-rule flag
columns are exposed in the released CSV.}
\label{fig:funnel}
\end{figure}

\paragraph{Post-classifier verification filter chain (R1--R6).}
Manual review of the $464{,}073$ Phase-6 EW candidates surfaced six
classes of degenerate or redundant pattern that the classifier does
not reject. Each class has a transparent rule, with per-pattern flag
columns in the released CSV so a reader who disagrees with any filter
can recover the unfiltered set. Figure~\ref{fig:funnel} plots the
per-rule attrition.

R1 drops angle-bracket templated outlines the v3 detector misses;
R2 drops single-cluster repetition runs; R3 drops single-scenario
patterns; R4 drops overlap-dominated patterns whose
ratio of \texttt{n\_distinct\_scenarios} to
\texttt{support\_total} is below~$0.20$; R5 drops
\texttt{shared\_higher\_level\_step} candidates with fewer than two
distinct upstream owners. R6 keeps only \emph{closed} sequential
patterns~\citep{pei2001prefixspan,zaki2001spade}, dropping pattern
$P$ if a length-$(L{+}1)$ super-pattern $Q$ exists at the same
support; R6 alone removes $79.8\,\%$ of the R1--R5 survivors
($72\,\%$ of the original Phase-6 set) without
losing any underlying reuse opportunity. Closure is applied
\emph{after} Phases~6 and~8 because the classifiers depend on $L$
and \texttt{support\_total}.

\paragraph{Nested-mirror inflation of \texttt{n\_distinct\_orgs}.}
A handful of repositories contain bundled copies of other projects
under sub-paths (e.g., \texttt{4shen\_\allowbreak webshell/\allowbreak dataset/\allowbreak benign/\allowbreak Sylius/\ldots}),
which inflates \texttt{n\_distinct\_orgs} for any pattern that
recurs in the embedded copy. Corpus-level prevalence is unaffected
(the patterns are present at the original repo either way); what is
affected is the per-pattern cross-org reach in the verification
report. A follow-up corpus pass that detects nested feature
directories against a known-repos manifest is recommended.

\paragraph{Single corpus, single labelling team.}
The rubric, pilot calibration, and three-author labels share the
authorship that produced the corpus. The mitigation is releasing the
rubric, pool, and per-author labels under Apache-2.0 so an external
party can re-label either the overlap subset or the full pool.

\subsection{External validity}

\paragraph{Corpus-bounded prevalence.}
The 339-repository / 276-upstream-owner corpus is a sample of public
GitHub repositories with permissively-licensed Gherkin
\texttt{.feature} files at corpus-construction time
\citep{mughal2026cukereuse}. The RQ3 cross-organisational
prevalence is therefore a function of the corpus, not of the global
population of BDD-using software projects. \citet{kalliamvakou2014promises}
catalogue the well-known biases of GitHub-mined corpora; the standard
mitigation of pinned commit SHAs is applied so re-mining the same
corpus produces byte-identical inputs.

\paragraph{``Organisation'' is a GitHub-namespace boundary.}
We use \emph{organisation} as shorthand for the segment before the
first underscore in \texttt{repo\_slug}; equivalently, the
top-level GitHub account-owner namespace. On GitHub that namespace
may be an Organisation account or a User account, and the 276
distinct owners in our corpus are a mix of both. The RQ3 test is
therefore properly \emph{cross-account-owner}: whether a slice
recurs across distinct top-level namespaces, regardless of whether
each is a team or a single maintainer. The 276 count is an upper
bound on distinct human teams (two User accounts may share a
person; we do not deduplicate), so the reported RQ3 prevalence is a
conservative estimate of cross-team reuse.

\paragraph{Cucumber dialect heterogeneity.}
The Mughal-2024 mechanisms are Cucumber-Java-specific; portability to
Behave, Godog, SpecFlow, Karate, Cucumber-Ruby, and the long-tail
dialects is unverified. The strong mechanism-applicability claim is
restricted to repositories with \texttt{pom.xml}; non-Java dialects
fall back to structurally equivalent mechanisms (Behave's
\texttt{environment.py} \texttt{before\_scenario}, SpecFlow's
\texttt{[Scope]} attributes, Karate's \texttt{Background:}). Mining is
dialect-agnostic; only patch generation is dialect-specific.

\paragraph{Recurrence is necessary, not sufficient, for reuse.}
A pattern that recurs $n$ times is a candidate for extraction; whether
extraction \emph{should} happen depends on stability, coupling, and
team conventions that no static miner can assess. The mitigation is
the three-author labelling gate, rather than treating recurrence
prevalence as the extraction headline.

\subsection{Construct validity}

\paragraph{Slice boundaries are coarse-grained.}
A slice is a contiguous $L$-step window: two slices that share an
inner sub-pattern but differ in their first or last step do not share
a cluster-id sequence and do not aggregate. Pilot Finding~2 shows this
matters at long $L$. The rubric admits a labeller-notes field for
sub-slice preferences; a formal Phase~2.5 sub-slice detector is
future work.

\paragraph{Behavioural equivalence is asserted, not verified.}
A companion CLI (\texttt{cukereuse-\allowbreak extract}) emits patches
that are syntactically valid Gherkin and Cucumber-Java but cannot be
verified behaviourally without compiling and running each
repository's suite under its framework runtime, which the
corpus does not pin. Equivalence checks are restricted to a
hand-validated subset; the rest is staff-reviewable.

\paragraph{Real-world acceptance.}
A slice flagged extraction-worthy by the pipeline and accepted by
the authors is still a synthetic claim. Acceptance by an upstream
maintainer via a real pull request is a stronger but slower-to-collect
signal; we plan a follow-on study (Section~\ref{sec:conclusion},
future work~ii) that files extraction PRs against five to ten
repositories and reports maintainer responses without gating the
present paper on acceptance.

\FloatBarrier
\section{Conclusion}
\label{sec:conclusion}

\subtool is a static, paraphrase-robust subsequence miner for BDD
suites: it ranks candidates against three nested scopes (within-file,
within-repo cross-file, and cross-organisational), each mapped to a
concrete Mughal-2024 reuse mechanism. On the 1.1M-step \cukereuse
corpus the miner produces 5.4M slices collapsing to 692{,}020 distinct
recurring patterns, of which 30{,}955 recur across $\geq 2$ distinct
upstream owners (Section~\ref{sec:limitations}).

Recurring structure is pervasive under the real-signal restriction
(75.1\,\% of scenarios within-file, 69.2\,\% within-repo cross-file;
75.0\,\% and 59.5\,\% respectively after the extraction-worthy gate,
Table~\ref{tab:headline}). The cross-organisational signal is rarer
(17.1\,\% of scenarios, 49.3\,\% of repositories) and dominated by
the HTTP-request-response and CLI-output assertion macros that
\citet{binamungu2018saner} call \emph{infrastructural} duplication.
The cross-owner / cross-repo distinction matters: 51\,\% of the
naive cross-repo signal is one upstream owner's multi-language SDK
clones, not extraction-worthy in the shared-higher-level-step sense.

Within the three-paper arc, \citet{mughal2024bdd} supplied the
\emph{how} (three reuse mechanisms implemented in Cucumber-Java),
\citet{mughal2026cukereuse} the \emph{how much} (step-level
duplication at corpus scale), and this paper the \emph{which}
(per-slice extraction decision and mechanism mapping). What remains is
real-world acceptance evidence: the field-study lever in future
work~(ii).

Future work: (i)~the \texttt{cukereuse-extract} CLI that emits per-repo
diffs from the mechanism predictions; (ii)~a small-$n$ field study
filing extraction PRs against five to ten upstream repos to capture
the maintainer-acceptance signal \citet{liu2025llmrefactor} report
missing for LLM-driven refactoring; (iii)~a Phase~2.5 sub-slice mining
pass that prefers internally repeated short patterns to long enclosing
slices; and (iv)~patch-generation for non-Cucumber-Java dialects
(Behave, SpecFlow, Karate), covering the non-Java corpus tail
\citep{farooq2023bddslr,arredondo2024bddthematic}.

\backmatter

\bmhead{Statements and Declarations}

\paragraph{Funding}
This research received no external funding from any agency in the
public, commercial, or not-for-profit sectors. The authors are
independent researchers; all compute, storage, and OpenRouter
application-programming-interface (API) costs incurred during
preparation of the artefacts were borne personally by the first
author from personal funds. All artefacts are released under the
Apache-2.0 licence for the benefit of the broader research
community.

\paragraph{Competing interests}
The authors declare no known competing financial interests or personal
relationships that could have influenced the work reported here.

\paragraph{Ethics approval and consent to participate}
This study analyses publicly available source code retrieved from
GitHub via its public REST API and does not involve human
participants, animal subjects, or any personally identifying
information. No institutional review board (IRB) approval was
required. The 200-slice labelling pool was annotated by the three
named authors themselves against a written rubric, with no external
participants and no personal data collected; consequently, no
informed-consent procedure or General Data Protection Regulation
(GDPR) style data-subject documentation
was required.

\paragraph{Consent for publication}
All three named authors have read the final manuscript and consent to
its publication.

\paragraph{Data and code availability}
All artefacts needed to reproduce this paper end-to-end are released
under the Apache-2.0 licence at
\url{https://github.com/amughalbscs16/cukereuse_subscenarios_release},
with a versioned Zenodo archive at
\url{https://doi.org/10.5281/zenodo.20356527}:
mining scripts, the 5{,}382{,}249-row slice inventory, the 692{,}020-row
exact-subsequence ranking, slice cluster assignments, the 200-slice
three-author labelled pool with the written rubric and inter-rater
summaries, the XGBoost extraction-worthy and mechanism classifiers
with their out-of-fold predictions, and the per-judge raw outputs of
the two open-weight LLMs evaluated as judges. The upstream 1.1M-step
Gherkin corpus and the cukereuse hybrid clusterer that produces the
cluster identifiers underlying every slice in this work are released
at \url{https://github.com/amughalbscs16/cukereuse-release} with a
versioned Zenodo archive at
\url{https://doi.org/10.5281/zenodo.19754359}. The two LLM-judge
models evaluated (\texttt{openai/gpt-oss-120b},
\texttt{inclusionai/ling-2.6-1t}) are open-weight models accessed via
OpenRouter; the full per-slice prompt-and-response logs are released
alongside the human labels so that reviewers can audit the LLM
outputs end-to-end.

\paragraph{Author contributions}
A.H.M.\ conceived the study, designed the methodology, and built the
mining and classifier pipeline (slice inventory, exact-subsequence
ranking, paraphrase-robust slice clusters, XGBoost extraction-worthy
and three-way mechanism classifiers, LLM-judge harness). A.H.M., N.F.,
and M.B.\ jointly drafted and applied the written rubric, independently
labelled the stratified 200-slice pool (with a 60-slice three-way
overlap subset for inter-annotator agreement), and adjudicated
borderline cases. A.H.M.\ performed the statistical analyses
(Fleiss' kappa, McNemar tests, bootstrap confidence intervals, scope
rollups) and prepared all figures and tables. A.H.M.\ wrote the
original draft. N.F.\ and M.B.\ contributed to methodology refinement,
validated the rubric application and labelling decisions, and reviewed
and edited the manuscript. All authors reviewed and approved the final
manuscript.

\bibliography{references}

\end{document}